\documentclass[fleqn,usenatbib]{mnras}

\usepackage{newtxtext,newtxmath}
\usepackage{changepage}
\usepackage{enumitem}
\usepackage{hyperref}
\PassOptionsToPackage{hyphens}{url}

\usepackage[T1]{fontenc}

\DeclareRobustCommand{\VAN}[3]{#2}
\let\VANthebibliography\thebibliography
\def\thebibliography{\DeclareRobustCommand{\VAN}[3]{##3}\VANthebibliography}

\usepackage{graphicx}	
\usepackage{amsmath}	
\usepackage{subcaption}
\usepackage{multirow}
\usepackage{subcaption}
\usepackage{mathtools}

\newcommand{\cmgr}{~cm$^{2}$\,g$^{-1}$}	
\newcommand{\geff}{{\gamma,\text{eff}}}

\newcommand{\asig}{\langle\Sigma_0\rangle}
\newcommand{\kb}{\text{k}_\text{B}\text{/b}}

\title[Thermalization of $\gamma$-rays]{The thermalization of $\gamma$-rays in radioactive expanding ejecta: A simple model and its application for Kilonovae and Ia SNe}

\author[Guttman et al.]{
Or Guttman\thanks{E-mail: or.guttman@weizmann.ac.il},
Ben Shenhar,
Arnab Sarkar
and Eli Waxman
\\
Department of Particle Physics \& Astrophysics, Weizmann Institute of Science, Rehovot 76100, Israel
}

\date{Accepted XXX. Received YYY; in original form ZZZ}

\pubyear{2024}

\begin{document}
\label{firstpage}
\pagerange{\pageref{firstpage}--\pageref{lastpage}}
\maketitle

\begin{abstract}

A semi-analytic approximation is derived for the time-dependent fraction $f_\gamma(t)$ of the energy deposited by radioactive decay $\gamma$-rays in a homologously expanding plasma of general structure. An analytic approximation is given for spherically symmetric plasma distributions. Applied to Kilonovae (KNe) associated with neutron stars mergers and Type Ia supernovae, our semi-analytic and analytic approximations reproduce, with a few percent and 10\% accuracy, respectively, the energy deposition rates, $\Dot{Q}_{\rm dep}$, obtained in numeric Monte Carlo calculations.
The time $t_\gamma$ beyond which $\gamma$-ray deposition is inefficient is determined by an effective frequency-independent $\gamma$-ray opacity $\kappa_\geff$, $t_\gamma = \sqrt{\kappa_\geff\langle\Sigma\rangle t^2}$, where $\langle\Sigma\rangle\propto t^{-2}$ is the average plasma column density. For $\beta$-decay dominated energy release, $\kappa_\geff$ is typically close to the effective Compton scattering opacity, $\kappa_\geff \approx 0.025$~cm$^{2}$\,g$^{-1}$ with a weak dependence on composition.
For KNe, $\kappa_\geff$ depends mainly on the initial electron fraction $Y_e$,
$\kappa_\geff \approx 0.03(0.05)$~cm$^{2}$\,g$^{-1}$ for $Y_e \gtrsim (\lesssim) 0.25$ 
(in contrast with earlier work that found $\kappa_\geff$ larger by 1--2 orders of magnitude for low $Y_e$), and is insensitive to the (large) nuclear physics uncertainties.
Determining $t_\gamma$ from observations will therefore measure the ejecta $\langle\Sigma\rangle t^2$, providing a stringent test of models. For $\langle\Sigma\rangle t^2=2\times10^{11}~{\rm g\,{cm}^{-2}\,s^2}$, a typical value expected for KNe, $t_\gamma\approx1$~d.
\end{abstract}




\section{Introduction}
\label{sec:intro}

Astronomical transients, such as Kilonovae (KNe) associated with neutron stars mergers (NSM) and Type Ia supernovae (Ia SNe), are powered by the radioactive decay of unstable isotopes in the rapidly expanding ejecta produced by the merger/explosion (\citealt{li_transient_1998,metzger_electromagnetic_2010} and \citealt{pankey_possible_1962,colgate_early_1969}; for recent reviews, see \citealt{metzger_kilonovae_2019,nakar_electromagnetic_2020} and \citealt{maoz_observational_2014}).
These radioactive decays release non-thermal particles: $\gamma$-rays, electrons and positrons, and possibly $\alpha$-particles and fission fragments that deposit part of their energy in the plasma. This energy deposition heats the plasma, driving the observed radiation. Understanding the efficiency of energy deposition is essential for the interpretation of the observed emission and for constraining the properties of the ejecta based on the observations.
This paper addresses the deposition efficiency of $\gamma$-rays.

In Type Ia supernovae, the heating mechanism is widely accepted. ${}^{56}$Ni is synthesized and then decays through the ${}^{56}$Ni $\rightarrow$ ${}^{56}$Co $\rightarrow$ ${}^{56}$Fe chain,
releasing $\gamma$-rays and positrons \citep{pankey_possible_1962,colgate_early_1969}.
Thus, the total radioactive energy rate produced is known, and the fraction of it that is deposited in the plasma ("thermalized") can be determined using the known spectra of decay products, with weak dependence on the exact ejecta composition.
Specifically, $t_0$ (denoted hereafter $t_\gamma$\footnote{As similar timescales exist for  $e^{\pm}$'s, $\alpha$'s and fission fragments.}), the time beyond which $\gamma$-rays begin to escape the ejecta without a significant energy deposition,
can be estimated using a simple frequency-independent (gray) effective opacity $\kappa_\geff \approx 0.025$\cmgr - the Compton scattering opacity weighted by the mean fractional energy loss in a single scattering event, hereafter referred to as the \textit{effective Compton scattering opacity} \citep{colgate_luminosity_1980,swartz_gamma-ray_1995}. This robust and analytic description of $t_\gamma$ allows to constrain the ejecta column density from observations \citep[e.g.][]{scalzo_type_2014,wygoda_type_2019-1}.

Kilonovae, on the other hand, pose a greater challenge. The neutron-richness of the matter that is ejected from NSM is thought to be sufficient for the production of elements heavier than iron by multiple and rapid neutron captures in a mechanism known as the r-process (\citealt{lattimer_black-hole-neutron-star_1974,symbalisty_neutron_1982,freiburghaus_r-process_1999}, see \citealt{fernandez_electromagnetic_2016,shibata_merger_2019,radice_dynamics_2020,rosswog_heavy_2024} for recent reviews). As the r-process ends when the neutron flux is depleted, a variety of isotopes that are neutron-rich and far from stability remain.
These isotopes decay mainly isobarically (along constant mass number paths) through $\beta$-minus decay (but under some conditions, also by $\alpha$ decay and spontaneous fission (SF)) back to the valley of stability.
This affects the heating rate in two ways. First, 
the total radioactive energy rate does not follow the simple exponential form of a single decay chain, but is rather closer to a power law, approximately $\sim t^{-1.3}$ (\citealt{li_transient_1998,metzger_electromagnetic_2010,hotokezaka_analytic_2017}; and similarly to \citealt{way_rate_1948}).
Second, predicting the deposition fraction of the injected radioactive power is more complex. The composition created by nucleosynthesis is sensitive to the ejecta conditions following the merger, which are poorly constrained, and to large nuclear physics uncertainties.
This results in poor constraints on the distribution of decaying isotopes, {on the spectra of emitted particles, and the stopping power of the ejecta.

The deposition of $\gamma$-rays in KNe was suggested to be sensitive to the ejecta composition \citep{hotokezaka_radioactive_2020,nakar_electromagnetic_2020,barnes_kilonovae_2021}, thereby having the potential for probing the nucleosynthesis conditions.
Neutron-richer conditions lead to the synthesis of heavier nuclei, with higher atomic numbers $Z$, which have the following properties:
(i) Higher-$Z$ elements have a larger cross-section for the absorption of low-energy (sub-MeV) $\gamma$-rays due to the photoelectric effect; 
(ii) Heavier nuclei have denser nuclear energy levels, and thus the $\gamma$-rays, emitted by isomeric transitions of excited nuclei following radioactive decay, are typically of lower energy.
These two properties acting together may thus increase the efficiency of $\gamma$-ray deposition, increasing $t_\gamma$ for low-$Y_e$ ejecta.

A robust (and preferably analytic) description of the $\gamma$-ray deposition in KNe, and particularly of $t_\gamma$, that will be an essential tool to derive constraints from observations, is currently lacking.
Past works on the subject either used approximate analytic estimates, which were not accurately verified by comparisons to the results of detailed $\gamma$-ray transport calculations \citep{metzger_electromagnetic_2010,barnes_radioactivity_2016, hotokezaka_radioactive_2016, hotokezaka_radioactive_2020}, or conducted detailed transport calculations with specific ejecta settings, that cannot be generalized to KNe of general properties (structure, composition) \citep{barnes_radioactivity_2016,barnes_kilonovae_2021}.
In between, approximate transport calculations used gray opacities derived from analytic works that were not verified by comparison to detailed calculations \citep[e.g.][]{wollaeger_impact_2018,bulla_critical_2023}.
Moreover, large differences exist in the estimated value of the effective $\gamma$-ray opacity $\kappa_\geff$. While some works use qualitative arguments and take $\kappa_\geff\approx0.025$\cmgr as in Ia SNe \citep[e.g.][]{metzger_electromagnetic_2010,hotokezaka_radioactive_2016}, others calculate an energy-weighted spectral average, that shows a strong dependence on $Y_e$ with values larger by one to two orders of magnitude  ($\approx0.07-3$\cmgr) at low $Y_e$, implying a significant dependence of $t_\gamma$ on $Y_e$ \citep{barnes_radioactivity_2016,hotokezaka_radioactive_2020,barnes_kilonovae_2021}.

In this paper, we derive an accurate semi-analytic approximation for $f_\gamma(t)$, the time-dependent fraction of the energy deposited by radioactive decay $\gamma$-rays, in a homologously expanding ejecta of general structure (i.e., general spatial dependence of density and atomic/isotopic composition).
An analytic approximation is given for spherically symmetric ejecta.
The resulting equations may be easily incorporated in calculations of light curves of KNe, Ia SNe, or other radioactively-powered transients, eliminating the need for time-consuming Monte Carlo (MC) calculations of $\gamma$-ray deposition, which is particularly challenging when a large range of model parameters needs to be explored.
For a similar analytic description of the deposition efficiency of $\beta$-decay electrons (and positrons) in NSM ejecta, see \citet{shenhar_analytic_2024}.

The validity of our approximations is demonstrated by comparisons to numeric MC calculations of $f_\gamma$ for common Ia SN ejecta models, and for a wide range of NSM ejecta density profiles and plasma parameters, that generously covers the expected range from binary NSM simulations \citep[e.g.][]{radice_binary_2018, nedora_numerical_2021, rosswog_heavy_2024}: initial electron fraction and entropy in the ranges $0.05\le Y_e\le0.45$, $1\le s_0[\kb]\le100$, and densities in the range $10^{-3}\le [\rho t^3]_{\text{KN}} \le10^3$. Here, $\rho t^3$ is normalized to the value obtained from the inferred mass and characteristic velocity of the ejecta of GW170817 \citep[e.g.][]{drout_light_2017, hotokezaka_neutron_2018,kasen_origin_2017, kasliwal_illuminating_2017, waxman_constraints_2018}, 
\begin{equation}
    \label{eq:rhoKN}
    [\rho t^3]_{\text{KN}} \equiv \frac{\rho t^3}{0.05M_\odot/\left(4\pi(0.2c)^3\right)}.
\end{equation}

The (time-dependent) atomic and isotopic composition of KN ejecta obtained for given $\{Y_e,s_0,\rho t^3\}$ are uncertain due to nuclear physics uncertainties. At low $Y_e$ this implies, for example, large (order of magnitude) uncertainties in the final abundance and radioactive energy production rate \citep{mumpower_impact_2016,zhu_modeling_2021,barnes_kilonovae_2021,kullmann_impact_2023,lund_influence_2023}. We show, by carrying out numeric calculations with different mass models and accounting for uncertainties in nuclear reactions' cross-sections, that $f_\gamma$ is nearly independent of the nuclear physics uncertainties.

One of the main results of our analysis is that the effective $\gamma$-ray opacity, $\kappa_\geff$, may differ widely from the (energy-release weighted) average of $\kappa_{\gamma,E}$, the opacity for $\gamma$-rays of energy $E$.
The reason for this may be easily understood as follows.
Consider the case of ejecta with uniform (space-independent) opacity $\kappa_{\gamma,E}$ and isotopic composition, producing radioactive decay $\gamma$-rays with a normalized (energy-release weighted) spectrum $\phi_\gamma$, 
\begin{equation}
\label{eq:gamma_spectrum}
    \phi_\gamma(E,t) = \Dot{Q}^{-1}_\gamma E\frac{d\Dot{N}_{\gamma}}{dE}(E,t), \quad \Dot{Q}_\gamma(t) = \int{ dE E\frac{d\Dot{N}_{\gamma}}{dE}(E,t)}.
\end{equation}
Here, $(d\Dot{N}_{\gamma}/dE)dE$ is the rate of $\gamma$-ray production at time $t$ in the energy interval $dE$ around $E$ and $\Dot{Q}_\gamma(t)$ is the total rate of $\gamma$-ray energy production. Assuming instantaneous energy deposition (valid for non-relativistic expansion) and denoting by $f_\gamma(E,t)$ the fraction of energy deposited by a $\gamma$-ray emitted at time $t$ with energy $E$, we have
\begin{equation}
    f_{\gamma}(t) = \int{dE \phi_\gamma(E,t)f_\gamma(E,t)}.
    \label{eq:f_gamma_all_energies}
\end{equation}
As the ejecta expands homologously, the column density falls off as $\Sigma\propto t^{-2}$. At late time, $t\gg t_\gamma$, the ejecta becomes optically thin to all $\gamma$-rays, $f_{\gamma}(E,t)\ll1$, and $ f_{\gamma}(E,t)$ is given by
\begin{equation}
    f_{\gamma}(E,t) = \kappa_{\gamma,E}\langle\Sigma\rangle.
    \label{eq:f_gamma_E_late}
\end{equation}
Here $\langle\Sigma\rangle\propto t^{-2}$ is the column density properly averaged over the location and direction of $\gamma$-ray emission (see \S~\ref{subsec:derivations}), and $\kappa_{\gamma,E}$ is defined as the fraction of energy lost to deposition by $\gamma$-rays of energy $E$ propagating through a unit column density $dX$,
\begin{equation}
\label{eq:k_def}
    \kappa_{\gamma,E}\equiv E^{-1}\left(\frac{dE}{dX}\right)_{\rm dep}(E).
\end{equation}
Combining eqs.~(\ref{eq:f_gamma_all_energies}) and~(\ref{eq:f_gamma_E_late}), we have
\begin{equation}
    f_{\gamma}(t) = t^{-2} \langle\kappa_{\gamma,E}\rangle \left(\langle\Sigma\rangle t^2\right),\quad  \langle\kappa_{\gamma,E}\rangle= \int{dE \phi_\gamma(E,t)\kappa_{\gamma,E}}.
    \label{eq:f_gamma_late}
\end{equation}
At late times the effective opacity that governs the deposition is indeed the (energy-release weighted) average of $\kappa_{\gamma,E}$, $\langle\kappa_{\gamma,E}\rangle$. 

This average may, however, lead to a large overestimate of the effective opacity at earlier times, particularly at $t\sim t_\gamma$, and hence to a large overestimate of $t_\gamma$ if $t_\gamma$ is estimated by $t_\gamma^{-2}\langle\kappa_{\gamma,E}\rangle (\langle\Sigma\rangle t^2)=1$. For SNe and KNe, a small fraction, $\phi_\gamma dE\ll1$, of the $\gamma$-ray energy production rate is carried by low-energy $\gamma$-rays/X-rays\footnote{$\gamma$-rays are produced by de-excitation of nuclei, at energies of $\sim$$10\text{'s keV}-\text{few MeV}$. X-rays are emitted by relaxation of atoms, at energies of $\sim$$1\text{keV}-10\text{'s keV}$.}. Due to the rapid increase of $\kappa_{\gamma,E}$ at low energy, it is often the case that at $t_\gamma$, when the higher energy $\gamma$-rays that carry most of the energy begin to escape the ejecta ($f_{\gamma}(E,t) \approx \kappa_{\gamma,E}\langle\Sigma\rangle<1$), the lower energy $\gamma$-rays/X-rays still deposit all their energy ($\kappa_{\gamma,E}\langle\Sigma\rangle>1$ and $f_{\gamma}(E,t) = \kappa_{\gamma,E}\langle\Sigma\rangle$ does not hold). Using eq.~(\ref{eq:f_gamma_late}) to estimate $f_\gamma$ would lead in this case to an overestimate of the deposition rate, due to the non-physical inclusion of low energy contributions with $f_{\gamma}(E,t) = \kappa_{\gamma,E}\langle\Sigma\rangle>1$ (which should be limited to $f_{\gamma}(E,t) =1$). This over-estimate may be large, even if only a small fraction of the energy is carried by low energy photons with $\kappa_{\gamma,E}\langle\Sigma\rangle\gg1$ at $t\sim t_\gamma$.

Earlier analyses of Ia SNe overcame this problem by neglecting the contribution of X-rays to the effective opacity $\langle\kappa_{\gamma,E}\rangle$ \citep[e.g.][]{swartz_gamma-ray_1995}. While this is a valid method for Ia SNe (as we show in \S~\ref{sec:type_ia}), it is not valid in general, and in particular, it fails for the case of KNe, due to the higher absorption cross-section at low $\gamma$-ray energies (see \S~\ref{sec:KNe}). 
Our treatment of the effective opacity properly accounts for both X-rays and $\gamma$-rays, despite the different energy scales, and does not require an arbitrary omission of low-energy contributions.

The paper is organized as follows. In \S~\ref{sec:gamma_model} we provide the
derivations of the semi-analytic and analytic approximations for $f_\gamma$ (\S~\ref{subsec:derivations}), describe the $\gamma$-ray energy loss processes and estimate $\kappa_\geff$ (\S~\ref{subsec:k_and k_eff}), and demonstrate the validity of the approximations by comparisons to MC calculations of $f_\gamma$ for well studied Ia SNe ejecta models (\S~\ref{sec:type_ia}). In \S~\ref{sec:KNe} we study $f_\gamma$ for KNe ejecta.
We derive $\kappa_\geff$ for KNe and show that it is insensitive to nuclear physics uncertainties. 
We then demonstrate the accuracy of the approximations for KN ejecta by comparisons to the results of detailed MC simulations, and compare our results to those of earlier studies.
Our results are summarized and discussed in \S~\ref{sec:summary}.

In comparing our analytic and semi-analytic approximations to the results of numeric MC calculations, we compare the results for both $f_\gamma$ and the total energy deposition rate,
\begin{equation}
   \label{eq:Q_dep}
    \Dot{Q}_{\text{dep}}(t) = f_\gamma(t)\Dot{Q}_{\gamma}(t) + \Dot{Q}_{\text{charged}}(t),
\end{equation}
where $\Dot{Q}_{\text{charged}}(t)$ is the rate of energy deposition by charged particles produced in radioactive decays: $e^{\pm}$, $\alpha$ and SF fragments. These particles lose energy much faster than $\gamma$-rays, hence the time beyond which their deposition efficiency drops well below unity is much longer than $t_\gamma$ \citep[at least by a factor of $\sim$10 for Ia SNe and KNe,][]{axelrod_late_1980,barnes_radioactivity_2016,waxman_late-time_2019,hotokezaka_radioactive_2020,barnes_kilonovae_2021,shenhar_analytic_2024}. Since charged particles dominate the energy deposition at times $t\gg t_\gamma$, obtaining accurate results for $f_\gamma$ at $t\sim t_\gamma$ is more important than at $t\gg t_\gamma$, particularly for KNe, where charged particles begin to dominate the energy deposition shortly after $t_\gamma$\footnotemark\citep[as $\Dot{Q}_{\text{charged}}(t)\sim\Dot{Q}_{\gamma}(t)$,][]{metzger_electromagnetic_2010,barnes_kilonovae_2021}. This affects some choices in our approximations. Finally, when comparing the approximations' results to detailed numeric calculations, we assume that charged particles deposit all their energy in the plasma instantaneously, which is a valid approximation for the time at which $\gamma$-ray deposition is significant.
\footnotetext{For Ia SNe, in contrast, $\Dot{Q}_{\gamma}(t)\gg\Dot{Q}_{\text{charged}}(t)$. However, for this low-$Z$ ejecta our approximations of $f_\gamma$ are highly accurate both at $t\sim t_\gamma$ and at later times (compare Ia SNe and KNe, in Figs. \ref{fig:typeIa_dep}, \ref{fig:kilonova_dep_frac}).}

\section{A Simple model for the $\gamma$-ray deposition}
\label{sec:gamma_model}

We consider the energy deposition of the non-thermal $\gamma$-rays and X-rays (referred to collectively as $\gamma$-rays) produced by radioactive decay in a homologously expanding ejecta of mass $M$ and characteristic velocity $\text{v}$, with a general density and atomic composition structure. We allow for multiple ejecta components with different atomic compositions and hence with different values of $\kappa_{\gamma,E}$ (i.e. allowing for spatially non-uniform $\kappa_{\gamma,E}$). We calculate $f_\gamma$ for a single component of radioactive material (i.e. assume a spatially uniform $\phi_\gamma$) with a density distribution that may differ from the total density of the ejecta. For the case of multiple radioactive components (with different isotopic compositions and $\phi_\gamma$ at different locations), $f_\gamma$ should be calculated separately for each component (and the energy deposition is obtained by a sum over all components, $\sum_i f_{i\gamma}\Dot{Q}_{i\gamma}$).

For homologous expansion, where the velocity $\textbf{v}$ of each fluid element is time independent, and its location is given by $\textbf{v}t$, the density of a fluid element with velocity $\textbf{v}$ at time $t$ is related to the density at some arbitrary reference time $t_0$ by $\rho(\textbf{v},t)t^3=\rho(\textbf{v},t_0)t_0^3$. The density distribution is thus determined by $\rho(\textbf{v},t_0)$ to which we refer as $\rho_0(\textbf{v})$. Similarly, $\rho_\text{0rad}(\textbf{v})$ is the density structure of the radioactively decaying material, which in general may differ from the total density $\rho_0(\textbf{v})$.

\subsection{The derivation of semi-analytic and analytic approximations}
\label{subsec:derivations}

The semi-analytic description is obtained using the following two approximations.
\begin{itemize}
    \item Neglecting the effects of relativistic expansion velocities.
    This implies instantaneous energy deposition (the energy is deposited at the same time the $\gamma$-ray is emitted, eq. \ref{eq:f_gamma_all_energies}), as well as neglecting the relativistic beaming of the emitted $\gamma$-rays, the red-shift of $\gamma$-rays with respect to the plasma rest-frame, and the temporal evolution of the ejecta during the propagation (from emission to absorption or escape) of a $\gamma$-ray through it.
    \item Approximating the fraction of energy deposited by a $\gamma$-ray of energy $E$ as $(1-e^{-\tau_{\gamma,E}})$ with an effective optical depth $\tau_{\gamma,E}=\int dl\rho \kappa_{\gamma,E}$, where the integral is over the $\gamma$-ray path along its emission direction (i.e. ignoring possible changes in propagation direction due to scattering) and neglecting the degradation of the $\gamma$-ray energy along the path.
    This approximation becomes exact at very early times when $\tau$ is large, and the energy is fully deposited, and at late times, when $\tau$ is small so that the energy loss is small and multiple scatterings do not occur, but underestimates the deposition at intermediate times, when $\tau\sim1$ and multiple scatterings and decrease of $E$ along the path (which increases $\kappa$) may contribute.
\end{itemize}
We show that the resulting semi-analytic description reproduces $\Dot{Q}_{\rm dep}$ obtained in detailed numeric calculations of both Ia SNe (which are non-relativistic) and KNe (with $\text{v}\sim0.2c$) with an accuracy of a few percent at all times
\citep[similar accuracy was obtained in earlier works on Ia SNe, employing gray transport with effective $\kappa$ obatined by neglecting the contribution of X-rays,][]{ambwani_gamma-ray_1988,swartz_gamma-ray_1995}.

Under these approximations, $f_\gamma(E,t)$ is given by (\citealt{jeffery_radioactive_1999}, identical to the gray transfer of \citealt{sutherland_models_1984,swartz_gamma-ray_1995})
\begin{equation}
    \label{eq:avg_f_gamma_angles}
    f_{\gamma}(E,t) = \int{d^3\text{v}\frac{\rho_{0\text{rad}}(\mathbf{v})t_0^3}{M_{\text{rad}}}\int{\frac{d\mathbf{\hat{\Omega}}}{4\pi} \left(1-e^{- \tau_\gamma(E,t;\mathbf{v},\mathbf{\hat{\Omega}})}\right)}}.
\end{equation}
Here, $M_\text{rad}$ is the mass of radioactive material (which may differ from $M$), and the effective optical depth along the path of a $\gamma$-ray emitted at $\textbf{r}=\textbf{v}t$ in the direction $\mathbf{\hat{\Omega}}$ is given by
\begin{equation}
    \label{eq:tau_g}
    \begin{split}
    \tau_\gamma(E,t;\mathbf{v},\mathbf{\hat{\Omega}}) &= 
    \int_0^\infty{du\, t\kappa_{\gamma,E}\rho(\textbf{r}=(\mathbf{v}+u\mathbf{\hat{\Omega}})t,t)
    } \\ &=
    t^{-2}\int_0^\infty{du  \kappa_{\gamma,E} \left[\rho_0(\mathbf{v}+u\mathbf{\hat{\Omega}}) t_0^3\right]}.
    \end{split}
\end{equation}
For uniform composition, $\kappa_{\gamma,E}$ is independent of location and
\begin{equation}
    \label{eq:tau_g_Sig}
     \tau_\gamma(E,t;\mathbf{v},\mathbf{\hat{\Omega}})=
     t^{-2} \kappa_{\gamma,E}[\Sigma_0(\mathbf{v},\mathbf{\hat{\Omega}})t_0^2],
\end{equation}
where $\Sigma_0(t/t_0)^{-2}$ is the column density at time $t$,
\begin{equation}
\label{eq:Sigma0}
    \Sigma_0(\mathbf{v},\mathbf{\hat{\Omega}}) = 
    \int_0^\infty{du t_0\rho_0(\mathbf{v}+u\mathbf{\hat{\Omega}}}).
\end{equation}
For multi-component ejecta, with different atomic compositions leading to different opacity at different locations, $\kappa_{\gamma,E}\Sigma_0$ is replaced by a sum of the contributions of the different components,
\begin{equation}
    \label{eq:tau_g_Sig_multi}
     \tau_\gamma(E,t;\mathbf{v},\mathbf{\hat{\Omega}})=
    t^{-2}\sum_i \kappa_{i\gamma,E}[\Sigma_{i0}(\mathbf{v},\mathbf{\hat{\Omega}})t_0^2],
\end{equation}
with 
\begin{equation}
    \label{eq:multi-Sigma}
    \Sigma_{i0}(\mathbf{v},\mathbf{\hat{\Omega}}) = 
    \int_0^\infty{du t_0\rho_{i0}(\mathbf{v}+u\mathbf{\hat{\Omega}}}),
\end{equation}
where $\rho_{i0}$ and $\kappa_{i\gamma}$ are the density and opacity of the $i$-th component.

At late time, when $\tau\ll1$, $f_\gamma(E,t)$ approaches 
\begin{equation}
    \label{eq:f_gE_late}
    f_\gamma(E,t)=(t/t_{\gamma,E})^{-2}
\end{equation}
with
\begin{equation}
    t^2_{\gamma,E}(t) = \sum_i\kappa_{i\gamma,E}(t) [\langle\Sigma_{i0}\rangle t_{0}^2]
    \label{eq:t_E_multi}
\end{equation}
and average column densities \citep[extension of][]{wygoda_type_2019-1}
\begin{equation}
    \label{eq:asig_multi_def}
    \langle\Sigma_{i0}\rangle t_0^2 = t_0^2\int{d^3\text{v}\frac{\rho_{\text{0rad}}(\mathbf{v})t_0^3}{M_{\text{rad}}}\int{\frac{d\mathbf{\hat{\Omega}}}{4\pi}}\mathrlap{\int_0^\infty}\qquad{du t_0\rho_{i0}(\mathbf{v}+u\mathbf{\hat{\Omega}})}}.
\end{equation}
We explicitly show in eq.~(\ref{eq:t_E_multi}) the dependence of $t_{\gamma,E}$ on time, due the possible time dependence of $\kappa_{i\gamma,E}$, which may arise from composition evolution with time. However, we note that for both Ia SNe and KNe, $\kappa_{i\gamma,E}$ and hence $t_{\gamma,E}$ are nearly independent of time (few percent variation for relevant energies). Finally, we note that the time dependence of $f_\gamma$ may deviate from $f_\gamma\propto t^{-2}$ also at late time, due to possible temporal evolution of $\phi_\gamma$.

\subsubsection{A simplified semi-analytic solution}
\label{sec:simplified semi}

As we show below, in \S~\ref{sec:type_ia} and in \S~\ref{sec:KNe}, the weak dependence of $\kappa_{\gamma,E}$ on composition for relevant $\gamma$-ray energies (those which carry most of the $\gamma$-ray energy) implies that a uniform $\kappa_{\gamma,E}$ is an excellent approximation for Ia SNe, and a very good approximation for KNe even under wide spatial variations of $Y_e$. This enables a further simplification of the semi-analytic approximation derived above. 

For a uniform $\kappa_{\gamma,E}$ the dependence of $f_{\gamma}(E,t)$ on $E$ and $t$ is only through the combination $\kappa_{\gamma,E}(t)t^{-2}$. $f_{\gamma}(E,t)$ is given in this case by a function of a single variable, 
\begin{equation}
\label{eq:gamma_E_t_semi}
f_{\gamma}(E,t) = f_{\xi}\left(\frac{t}{t_{\gamma,E}}\right), 
\end{equation}
where
\begin{equation}
\label{eq:fxi}
f_{\xi}(\xi) \equiv \int{d^3\text{v}\frac{\rho_{\text{0rad}}(\mathbf{v})t_0^3}{M_{\text{rad}}}\int{\frac{d\mathbf{\hat{\Omega}}}{4\pi}\left(1-e^{-\xi^{-2}\Sigma_0(\mathbf{v},\mathbf{\hat{\Omega}})/\asig}\right)}},    
\end{equation}
with 
\begin{equation}
    \label{eq:asig_def}
    \begin{split}
    \asig t_0^2 & = t_0^2\int{d^3\text{v}\frac{\rho_{\text{0rad}}(\mathbf{v})t_0^3}{M_{\text{rad}}}\int{\frac{d\mathbf{\hat{\Omega}}}{4\pi}}\int_0^\infty{du t_0\rho_0(\mathbf{v}+u\mathbf{\hat{\Omega}})}}\\
    & = f_\Sigma \frac{M}{4\pi \text{v}^2}.
    \end{split}
\end{equation}
Here, $f_\Sigma$ is a dimensionless factor (of order unity for spherically-symmetric ejecta). 

For uniform $\kappa_{\gamma,E}$, eqs.~(\ref{eq:gamma_E_t_semi})-(\ref{eq:asig_def}) with $t_{\gamma,E}(t)= \sqrt{\kappa_{\gamma,E}(t) \asig t_0^2}$ (eq.~\ref{eq:t_E_multi}), yield results identical to those of eqs.~(\ref{eq:avg_f_gamma_angles})-(\ref{eq:Sigma0}). For multi-component ejecta, eqs.~(\ref{eq:gamma_E_t_semi})-(\ref{eq:asig_def}) with $t_{\gamma,E}(t)$ given by eq.~(\ref{eq:t_E_multi}) provide an approximation to eqs.~(\ref{eq:avg_f_gamma_angles})-(\ref{eq:multi-Sigma}), corresponding to a uniform opacity 
\begin{equation}
    \label{eq:k_av}
    \Bar{\kappa}_{\gamma,E}(t)\equiv\asig^{-1}\sum_i\kappa_{i\gamma,E}(t)\langle\Sigma_{i0}\rangle,
\end{equation}
where $\langle\Sigma_0\rangle=\sum_i\langle\Sigma_{i0}\rangle$. This is evident by noting that 
\begin{equation}
    t^2_{\gamma,E}(t) = \Bar{\kappa}_{\gamma,E}(t) [\langle\Sigma_{0}\rangle t_{0}^2].
    \label{eq:t_E_av}
\end{equation}

The inaccuracy introduced by replacing the spatially dependent $\kappa_{\gamma,E}$ with the component average $\Bar{\kappa}_{\gamma,E}$ may be estimated as follows. Opacity variations $\delta\kappa$ introduce a correction $(\delta\kappa/\Bar{\kappa})\tau$ to $\tau$, which in turn introduces a correction $(\delta\kappa/\Bar{\kappa})\tau$ to the integrand $1-\exp(-\tau)$ of eq.~(\ref{eq:avg_f_gamma_angles}) for $\tau\le1$ (and smaller for $\tau>1$). Since at $t\sim t_\gamma$ we have $\tau\sim1$ for relevant $\gamma$-ray energies and a significant fraction of the emission positions and directions, the fractional correction to $f_\gamma(t)$ is expected to be $\lesssim \delta\kappa/\Bar{\kappa}$. As we show below, the effective opacity for KNe varies between 0.03 and 0.05\cmgr, implying that the introduced fractional inaccuracy is $\approx20\%$ for KNe (and negligible for Ia SNe, as discussed in \S~\ref{sec:type_ia}). This conclusion is supported by the results of numeric calculations presented in \S~\ref{sec:KNe}.

\subsubsection{Analytic approximation for spherically symmetric ejecta}
\label{sec:Analytic}

\citet{sharon_-ray_2020} have shown that for spherically symmetric density profiles appropriate for nickel-powered core-collapse supernovae $f_\gamma(t)$ is well approximated by analytic functions of the form
\begin{equation}
\label{eq:f_geff_general}
    f_{\gamma\rm, an.}(t)= \left[1+(t/t_\gamma)^n\right]^{-2/n}
\end{equation}
with $n \approx 2-4$, interpolating between $f=1$ at early time and $f\propto t^{-2}$ at late time. We show in \S~\ref{sec:type_ia} and \S~\ref{sec:KNe} that this functional form, with $t_\gamma$ and $n$ defined below, provides a good approximation also for the (spherically symmetric) Ia SN and KN cases, respectively, yielding $\sim$10\% accuracy in $\Dot{Q}_{\rm dep}$ at all times. 

To determine $t_\gamma$, the time at which the deposition efficiency drops significantly below unity, we approximate $f_\gamma(E,t)$ by interpolation between the early and late time behaviors, $f_\gamma(E,t)=1-\exp(-t_{\gamma,E}^2(t)/t^2)$, and define $t_\gamma$ as the time at which $f_\gamma(t)$ drops to $1-e^{-1}$. This defines $t_\gamma$ implicitly,
\begin{equation}
	\label{eq:t_geff_integral_def_2}
	\int{dE \phi_\gamma(E,t_\gamma) \left( 1 - e^{-t_{\gamma,E}^2(t_\gamma)/t_\gamma^2}\right)}=1-e^{-1}.
\end{equation}
$t_\gamma$ may be used to define an effective frequency independent $\gamma$-ray opacity, $\kappa_\geff$, by
\begin{equation}
	\label{eq:t_geff_def}
	t_\gamma \equiv \sqrt{\kappa_\geff \asig t_0^2},
\end{equation}
which is equivalent to defining $\kappa_\geff$ implicitly by
\begin{equation}
	\label{eq:k_geff_integral_def_2}
\int{dE \phi_\gamma\left(E,t=\sqrt{\kappa_\geff \asig t_0^2}\right) \left( 1 - e^{-\Bar{\kappa}_{\gamma,E}/\kappa_\geff}\right)}=1-e^{-1}.
\end{equation}

$\kappa_\geff$ defined by eq.~(\ref{eq:k_geff_integral_def_2}) is a function of the $\gamma$-ray spectrum, $\phi_\gamma$, of the opacities, $\kappa_{\gamma,E}$ and of the column density, $\asig$. However, as we show below, the dependence of $\kappa_\geff$ on $\asig$ is very weak, i.e. $\kappa_\geff$ depends mainly on the atomic and isotopic compositions. This implies, in particular, that the dependence of $t_\gamma$ on $\asig$ is approximately given by $t_\gamma\propto\asig^{1/2}$.
For the analysis of results below, it is useful to define a time-dependent effective opacity, $\kappa^t_\geff(t)$, by
\begin{equation}
\label{eq:kgeff_t_definition}
\int{dE \phi_\gamma(E,t) \left(1-e^{-\Bar{\kappa}_{\gamma,E}(t)/\kappa^t_\geff(t)}\right)}=1-e^{-1}.
\end{equation}
$\kappa^t_\geff(t)$ is a function of the $\gamma$-ray spectrum and opacities at time $t$ only, and is strictly independent of $\asig$. It may be considered as the effective opacity for an ejecta with $\asig$ for which $t_\gamma=t$.
The time dependence of $\kappa^t_\geff(t)$ serves as a measure of effect of the variation with time of $\phi_\gamma$ and of $\kappa_{\gamma,E}$.

$n$ is chosen such that $f_{\gamma\rm, an.}(t=t_\gamma)=f_{\xi}(\xi=1)$,
\begin{equation}
    n = -\frac{2 \ln{2}}{\ln{f_{\xi}(1)}}.
    \label{eq:n_interpolation}
\end{equation}
This choice provides a good approximation for $f_\gamma$ at $t\sim t_\gamma$, i.e., at the transition from efficient to inefficient deposition, for cases where $\phi_\gamma$ is dominated at this time by some characteristic energy $E$ (or more generally, characteristic $\kappa_{\gamma,E}$), and hence $t_\gamma$ is approximately given by $t_{\gamma,E}$ of this energy, and $f_\gamma(t_\gamma)\approx f_\gamma(E,t_\gamma) \approx f_{\xi}(1)$. 

$\asig t_0^2$ and $n$ are given for a few spherical density profiles of KNe and Ia SNe ejecta in Table \ref{table:interpolation}. A code for the numerical calculation of the semi-analytic expressions, eqs. \eqref{eq:avg_f_gamma_angles}-\eqref{eq:t_E_av}, including $f_\gamma(E,t)$, $f_\xi$, and $\asig t_0^2$, as well as of $n$ used in the analytic approximation (eq. \ref{eq:n_interpolation}), for multi-component ejecta with general structures $\rho_0(\textbf{v})$ and $\rho_{0\rm rad}(\textbf{v})$ is available at \url{https://github.com/or-guttman/gamma-ray-deposition}.

\begin{table}
 \caption[Caption for LOF]{\label{table:interpolation}The average column density $\langle\Sigma\rangle t^2$ (eq. \ref{eq:asig_def}) and the interpolation parameter $n$ of the analytic approximation (eq. \ref{eq:n_interpolation}), computed for spherical density profiles of KNe and Ia SNe ejecta.
 For KNe, all mass is assumed radioactive, $\rho_\text{rad}=\rho$, and for Ia SNe $\rho_\text{rad}$ is set by the initial ${}^{56}$Ni mass fraction of the models.
$\langle\Sigma\rangle t^2$ obtained for the uniform density shell model and the model of \citet{waxman_constraints_2018} is approximately equal to the radial column density $t^2\int_0^\infty{d\text{v}t~\rho(\text{v},t)}$, that equals $\rho t^3\Delta \text{v}$ ($\Delta \text{v} = \text{v}_{\rm max} - \text{v}_{\rm min}$) and $\frac{M}{4\pi (1+2\alpha)\text{v}^2_\text{M}}$, respectively. These radial column densities and the values of $n$ given provide good approximations for any parameters of these models (up to a order unity factor for $\langle \Sigma \rangle t^2$ and up to few 10's\% for $n$).
 For a radioactive sphere of uniform density truncated at $\text{v}=\text{u}$, $\langle\Sigma\rangle t^2$ is $\frac{3}{4}\rho t^3\text{u}$, and $n=2.4$.}
\centering
\begin{tabular}{|p{1em}|p{13em}|p{11em}|p{0.7em}|} 
 \hline
 Type & Model & $\langle\Sigma\rangle t^2~[{\rm g\,{cm}^{-2}\,s^2}]$ & $n$ \\ 
  \hline
 \multirow{3}{2em}{KNe} 
 & Shell of uniform density, \newline from $\text{v}_{\rm min}=0.1\text{c}$ to $\text{v}_{\rm max}=0.3\text{c}$\vadjust{\vspace{-0.5em}}\newline &  $2.2\times10^{11}\frac{M}{0.05M_\odot}$ & 2.4 \\ 
 & \citet{waxman_constraints_2018} - eq. 1, \newline with $v_{\rm M} = 0.15\text{c}$ and $\alpha = 0.7$ \newline as obtained for AT2017gfo \newline ($\rho(\text{v})$ is truncated at $\text{v}=\text{c}$) & $1.7\times10^{11}\frac{M}{0.05M_\odot}$ & 1.9 \\ 
 & \citet{barnes_effect_2013} - eq. 5, \newline with characteristic velocity $\text{v}_{\rm char}$ & $3.2\times10^{11}\frac{M}{0.05M_\odot}\left(\frac{\text{v}_{\rm char}}{0.2\text{c}}\right)^{-2}$ & 2.1 \\ 
 \hline
 \multirow{2}{2em}{Ia SNe} & \citet{blondin_standart_2022}, toy06 & $4.4\times10^{14}$ & 2.3 \\
  & \citet{blondin_standart_2022}, toy01 & $7.2\times10^{14}$ & 2.8 \\
 \hline
\end{tabular}
\end{table}

\subsection{$\gamma$-ray energy loss and the value of $\kappa_\geff$}
\label{subsec:k_and k_eff}

\subsubsection{$\gamma$-ray energy loss}
\label{subsec:thermalization_opacities}

$\gamma$-rays lose their energy while traveling through matter via three dominant mechanisms: photoelectric absorption (PE), Compton scattering, and pair-production (PP) \citep[see e.g.][for reviews]{shultis_radiation_2000,longair_high_2011}.
For our semi-analytic and analytic calculations, we assume that all the energy lost by a $\gamma$-ray in PE and Compton scattering interactions is instantaneously deposited in the plasma, while for PP interactions, the kinetic energy of the $e^{-}e^{+}$ pair is instantaneously deposited. This implies that $\kappa_{\gamma,E}$ (eq. \ref{eq:k_def}) is given by
\begin{equation}
    \label{eq:absorption_opacity}
    \kappa_{\gamma,E} = \kappa_{\textrm{PE}}(E) + \Bigl\langle\frac{\Delta E}{E}\Bigr\rangle\kappa_{\textrm{C}}(E) + \left(1-\frac{1.022\text{MeV}}{E}\right)\kappa_{\textrm{PP}}(E),
\end{equation}
where $\kappa_{\textrm{PE}}$, $\kappa_{\textrm{C}}$ and $\kappa_{\textrm{PP}}$ are the interaction opacities of PE, Compton scattering and PP, respectively \citep[taken from XCOM][]{berger_xcom_2010}, $\langle\frac{\Delta E}{E}\rangle$ is the fraction of energy transferred to an electron in a single Compton scattering averaged over scattering angles \citep[e.g.][]{sutherland_models_1984,swartz_gamma-ray_1995}, and $1-1.022\text{MeV}/E$ is the fraction of energy converted to kinetic energy of the $e^{-}e^{+}$ pair produced in PP interactions. This approximation is valid since the energy loss time of secondary electrons and positrons is short at $t\sim t_\gamma$ (see end of \S~\ref{sec:intro}), and since secondary low-energy fluorescence and Bremsstrahlung X-rays carry only a small fraction of the energy lost by the primary $\gamma$-ray and their interaction cross section is much higher than that of the primary $\gamma$-ray\footnote{While this may not hold for fluorescence X-rays emitted following incoming $\gamma$-rays of low-energies ($\lesssim$100 keV) in high-Z materials, leading to an overestimation of the deposition rate, only a small fraction of $\gamma$-ray energy is emitted at this energy range for typical $\beta$-emitters, and specifically for those present in KNe.}. 
We neglect the possible energy deposition by $\gamma$-rays produced in $e^{-}e^{+}$ annihilation (of $e^{+}$ produced by PP), similarly to neglecting multiple Compton scatterings and energy degradation along the propagation path (see \S~\ref{subsec:derivations}), so that $\kappa_{\gamma,E}\Sigma$ correctly reproduces the energy deposition fraction at late time, when $\kappa_{\gamma,E}\Sigma$~$\ll$ 1.
$\kappa_{\gamma,E}(t)$ for an atomic composition is obtained by the mass-weighted sum of the atomic $\kappa_{\gamma,E}$.

\begin{figure}
\centering
 \includegraphics[width=\columnwidth]{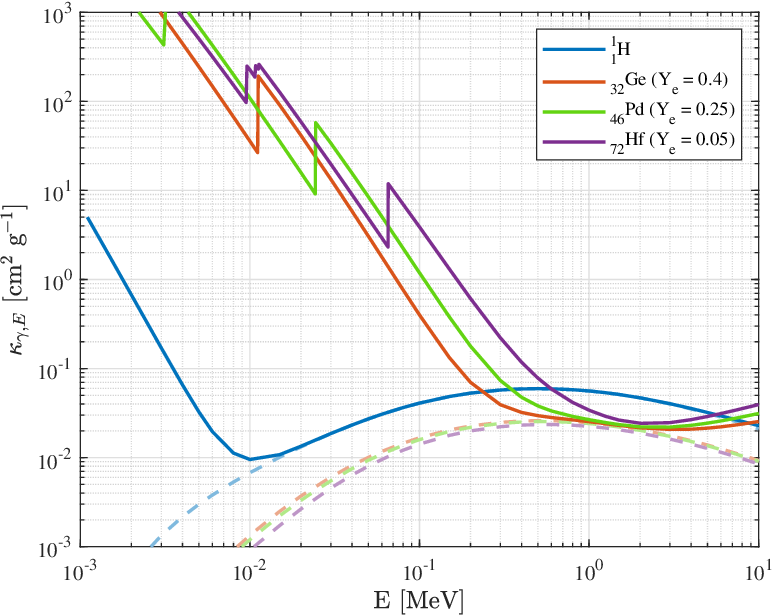} 
 \caption{The $\gamma$-ray opacity as a function of energy $\kappa_{\gamma,E}$ (eq. \ref{eq:absorption_opacity}), in solid lines, for  various elements: ${}^{1}_{1}$H (blue), ${}_{32}$Ge (orange), ${}_{46}$Pd (green) and ${}_{72}$Hf (purple). In dashed lines, we show the energy loss opacity due to Compton scattering $\langle\frac{\Delta E}{E}\rangle\kappa_{\textrm{C}}(E)$ (see text). This opacity is nearly energy-independent near 1~MeV and nearly material-independent as $\kappa_{\textrm{C}}\propto Z/A$ (thus ${}^{1}_{1}$H is an outlier). The opacities of $\{{}_{72}\text{Hf},{}_{46}\text{Pd},{}_{32}\text{Ge}\}$ represent well $\kappa_{\gamma,E}$ in NSM ejecta with initial $Y_e = \{0.05, 0.25, 0.4\}$ and with $[\rho t^3]_{\text{KN}}=1$ (eq.~\ref{eq:rhoKN}), $s_0 = 20~\kb$ (see results for the effective $Z$ of the ejecta in Fig. \ref{fig:kilonova_kappa_eff_examination}).}
 \label{fig:kappa_gamma_general}
\end{figure}

Fig. \ref{fig:kappa_gamma_general} shows $\kappa_{\gamma,E}$ for various atoms. Two important properties should be noted:
\begin{enumerate}
    \item For $\sim$1 MeV photons, the energy loss is dominated by Compton scattering, and thus only weakly dependent on energy and $Z$ (as $\kappa_{\textrm{C}}\propto Z/A$, where $A$ is the mass number, and $Z/A$ is roughly constant for all nuclei near the valley of stability, except ${}^{1}_{1}$H).
    \item PE is the dominant energy loss mechanism at low energies. Typically, the PE absorption is larger for higher atomic numbers and lower energies. Thus, as $Z$ increases, a larger energy range at $\lesssim$1 MeV is dominated by PE. However, at $\sim$1 MeV PE is at most comparable to Compton scattering. As a very crude approximation, at energies larger than the K-shell binding energy $\kappa_{\text{PE}} \sim (Z/E)^{2.5}$ \citep[similar rough scaling is given by][]{shultis_radiation_2000}.
\end{enumerate}

\subsubsection{The typical value of $\kappa_\geff$}
\label{subsubsec:eff_opacity}

$\beta$-decaying isotopes tend to emit most of their $\gamma$-ray energy as $\sim$1~MeV photons \citep[e.g.][]{brown_endfb-viii0_2018}. Accordingly, in an ejecta dominated by $\beta$-decay energy release, $\phi_\gamma$ is typically dominated by $\sim$1~MeV $\gamma$-rays.
Hence, eq.~\eqref{eq:kgeff_t_definition} gives $\kappa^t_\geff(t) \approx \kappa_{\gamma,1\text{MeV}}(t)$.

As described in \S~\ref{subsec:thermalization_opacities}, Compton scattering is the dominant energy loss mechanism at $\sim$1~MeV photon energy, while PE contributes a comparable (energy loss) opacity only for high-$Z$ ejecta. Thus, $\kappa^t_\geff(t)$ is nearly composition and time independent, and $\kappa_\geff$ is nearly composition and column density ($\langle\Sigma\rangle t^2$) independent: $\kappa_\geff$ is typically close to the effective Compton opacity $\approx 0.025$\cmgr for low-$Z$ ejecta, and is larger by (only) a factor of $\sim$2 for high-$Z$ ejecta.
Notably, this holds even if a significant fraction (few 10's of \%) of the $\gamma$-ray energy is emitted as low-energy $\gamma$-rays, with opacities higher by few orders of magnitude than those of $\sim1$~MeV photons. This is because the contribution of these low-energy photons to the effective opacity is exponentially suppressed in eq.~(\ref{eq:kgeff_t_definition}), relative to the contribution of the $\gamma$-rays that carry most of the energy.
Thus, $\kappa_\geff \gg 0.025-0.05$\cmgr would be obtained only for very soft $\gamma$-ray spectra that are atypical in general for $\beta$-decaying isotopes (see Appendix~\ref{ap:g-spec-large-kappa} for an expanded discussion).

\subsection{A Test Case: Type Ia supernovae}
\label{sec:type_ia}
\begin{figure}
    \centering
    \includegraphics[width=1\columnwidth]{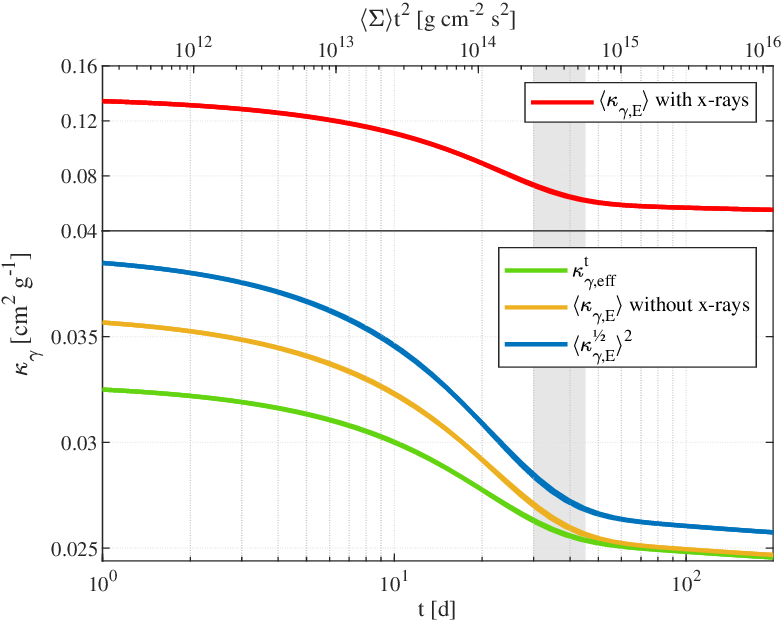}
    \caption{The results of different definitions of an effective $\gamma$-ray opacity as a function of time, for Ia SNe, assuming a pure initial ${}^{56}$Ni composition: $\kappa^t_\geff(t)$ (eq. \ref{eq:kgeff_t_definition}, green), the effective opacity that would be obtained from eq.~(\ref{eq:k_geff_integral_def_2}) for an ejecta with column density $\langle\Sigma\rangle t^2$ for which $t_\gamma=t$ (the value of $\langle\Sigma\rangle t^2$ for which $t_\gamma=t$ is shown on the top x-axis); $\langle\kappa_{\gamma,E}\rangle$, the energy-weighted average opacity over $\gamma$-rays and X-rays (eq. \ref{eq:f_gamma_late}, red); the energy-weighted average opacity excluding X-rays (gold); and $\langle\kappa^{1/2}_{\gamma,E}\rangle^2$, the opacity corresponding to defining $t_\gamma$ as an average over $t_{\gamma,E}$ (inclduing $\gamma$-rays and X-rays; blue, see text). The shaded gray region shows the relevant range of $t_\gamma$, roughly 30-45 d \citep[according to observations, ][]{wygoda_type_2019-1}.}
    \label{fig:typeIa_opacity}
\end{figure}

Here we apply our semi-analytic and analytic approximations for the $\gamma$-ray deposition to the case of Ia SNe. We show that: (i) Our approach naturally reproduces the effective $\gamma$-ray opacity commonly used in the literature, without the need to discard the contribution of low-energy fluorescence X-rays in an ad-hoc manner; (ii) The results of MC simulations of $\gamma$-ray deposition agree with our semi-analytic and analytic approximations for $f_\gamma$ to a few percent level.

For simplicity, we model the Ia SN ejecta with an initial composition of pure ${}^{56}$Ni, since the $\gamma$-ray deposition is insensitive to the exact composition (see bottom panel of Fig. \ref{fig:typeIa_dep}).
This is because the ejecta is composed of $Z\lesssim30$ elements, and most of the $\gamma$-ray energy is emitted as $\sim$1 MeV photons. Thus, the energy loss opacity is almost purely determined by Compton scattering, which is nearly composition-independent (Fig. \ref{fig:kappa_gamma_general}).
The $\gamma$-ray lines, X-ray lines and half-lives are taken from ENDF/B-VIII.0 \citep{brown_endfb-viii0_2018}.

 Fig. \ref{fig:typeIa_opacity} shows the time-dependent effective $\gamma$-ray opacity of Ia SNe $\kappa^t_\geff(t)$ as defined by eq.~\eqref{eq:kgeff_t_definition} (the effective opacity that would have been obtained from eq. \ref{eq:k_geff_integral_def_2} for an ejecta with column density $\langle\Sigma\rangle t^2$ for which $t_\gamma=t$). At the range of observed $t_\gamma$ for Ia SNe (gray region), $\kappa_\geff \approx 0.026$\cmgr, in agreement with the values used in the literature: $(0.05-0.06)Y_e\text{\cmgr}\approx 0.025-0.03\text{\cmgr}$, where $Y_e\approx 0.5$ for Ia SNe \citep{colgate_luminosity_1980,sutherland_models_1984,ambwani_gamma-ray_1988,swartz_gamma-ray_1995}\footnotemark.
\footnotetext{\citet{weaver_type_1980} used a larger opacity as they neglected non-radial column densities \citep[see][in the review of KEPLER]{blondin_standart_2022}. \citet{wilk_solving_2019} needed larger opacity to describe the \textit{local} energy deposition at early times \citep[due to multiple Compton scatterings,][]{swartz_gamma-ray_1995}, but as seen in their figure 7, the \textit{total} energy deposition at early times is insensitive to the exact opacity used.}

The figure also shows the energy-weighted average opacity over the $\gamma$-ray (and  X-ray) spectrum, $\langle\kappa_{\gamma,E}\rangle$ (eq. \ref{eq:f_gamma_late}), which highly overestimates $\kappa_\geff$. The largest overestimate, by a factor of $\sim$4, occurs when ${}^{56}$Ni is the dominant heating source due to its larger X-ray emission (Fig. \ref{fig:typeIa_dep}).
As explained in \S~\ref{sec:intro}, this is the result of overestimating the deposition of soft $\gamma$-rays, for which the PE opacity is very large (compared to the Compton opacity), even as nearly all the energy is emitted at energies at which Compton scattering dominates, and the energy fraction of X-rays is very small $\lesssim0.1\%$.
\citet{swartz_gamma-ray_1995} computed the effective opacity using $\langle\kappa_{\gamma,E}\rangle$, while discarding the X-rays and PE.
Indeed, $\langle\kappa_{\gamma,E}\rangle$ calculated while discarding the X-rays (but not PE) as shown in Fig. \ref{fig:typeIa_opacity} is similar to $\kappa_\geff$, with a slight overestimation when ${}^{56}$Ni dominates, due to its softer $\gamma$-ray (not X-ray) emission.
This overestimation shows that even when X-rays are discarded, the effect of PE on the opacity of low-energy $\gamma$-rays will cause $\langle\kappa_{\gamma,E}\rangle$ to overestimate the true effective opacity.
The separation between X-rays and $\gamma$-rays is purely based on the emission mechanism, $\gamma$-rays of sufficient low-energy will lead to overestimation much like X-rays.  
Thus, neglecting the X-rays will not in general yield an accurate result, as in general $\gamma$-rays may suffer from larger PE absorption (due to softer spectrum or higher $Z$ composition) or X-rays may carry a larger fraction of the energy.
In contrast, our definition of $\kappa_\geff$ yields the correct result without discarding an interaction mechanism or some of the $\gamma$-ray spectrum.
Finally, the figure also shows that $\langle\kappa^{1/2}_{\gamma,E}\rangle^2$, corresponding to an estimate of $t_\gamma=\langle t_{\gamma,E}\rangle$ instead of $t_\gamma=\sqrt{\langle t^2_{\gamma,E}\rangle}$ corresponding to $\langle\kappa_{\gamma,E}\rangle$ (recall $t_{\gamma,E}\propto \kappa_{\gamma,E}^{1/2}$), provides a good approximation for $\kappa_\geff$ without omitting the X-rays.

\begin{figure}
    \centering
    \includegraphics[width=1\columnwidth]{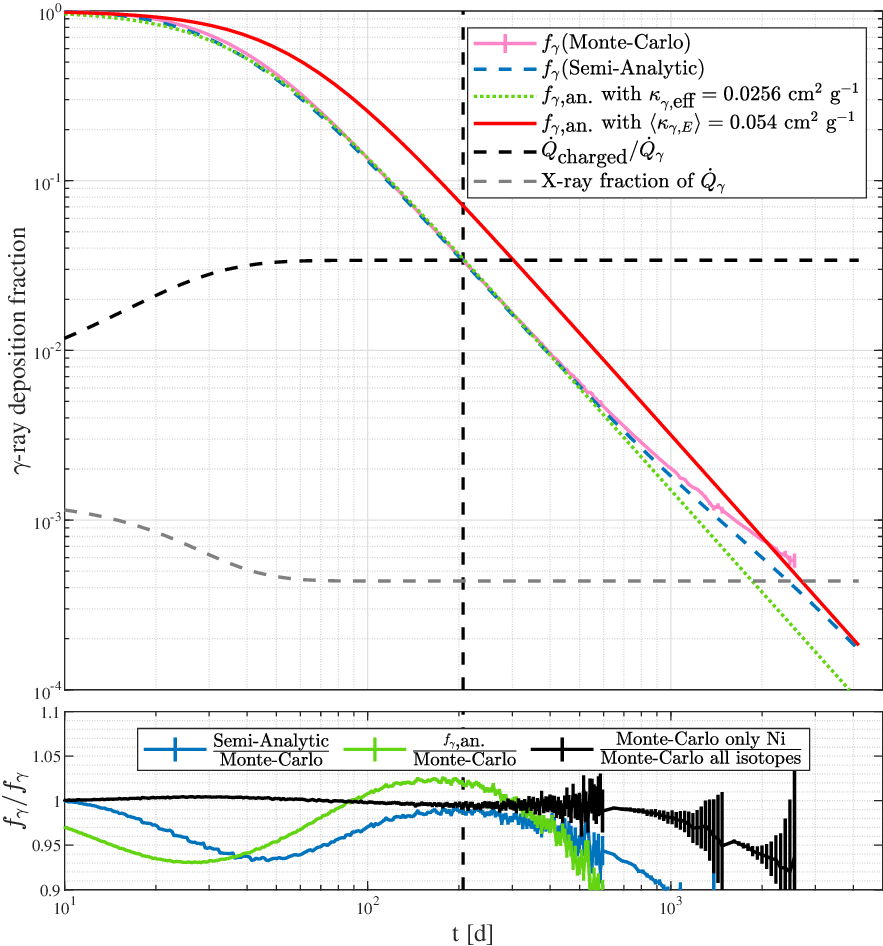}
    \caption{The $\gamma$-ray deposition fraction of an Ia SN toy model, with pure initial ${}^{56}$Ni composition and the density profile of toy06 \citep[][Table \ref{table:interpolation}]{blondin_standart_2022}.
    \textit{Top}: In pink, the MC results, computed using URILIGHT (error bars are due to statistical uncertainty); In blue, the semi-analytic solution, eqs.~(\ref{eq:f_gamma_all_energies}), (\ref{eq:gamma_E_t_semi})-(\ref{eq:t_E_av}); In green, the analytic approximation, $f_{\gamma\rm, an.}$, eq. \eqref{eq:f_geff_general} with $t_\gamma = 38.8$ d corresponding to (eq. \ref{eq:t_geff_def}) $\kappa_\geff=0.0256$\cmgr obtained from eq. \eqref{eq:k_geff_integral_def_2}, and $n=2.26$ obtained from eq. \eqref{eq:n_interpolation} (Table \ref{table:interpolation}); In red, $f_{\gamma\rm, an.}$, eq. \eqref{eq:f_geff_general}, with $t_\gamma = 56$ d corresponding to $\kappa_\geff = \langle\kappa_{\gamma,E}\rangle=0.054$\cmgr. 
    The horizontal black and gray dashed lines show $\dot{Q}_{\text{charged}}/\dot{Q}_{\gamma}$ and the fraction of photon energy carried by X-rays, respectively. The vertical black dashed line denotes the time beyond which $\dot{Q}_{\text{charged}}$ dominates $\Dot{Q}_{\text{dep}}$.
    When $f_\gamma$ becomes similar to the X-ray fraction, its slope changes as the X-rays continue to deposit all their energy in the plasma. At very late time, $f_\gamma$ approaches the asymptotic $f_\gamma\approx\langle\kappa_{\gamma,E}\rangle\langle\Sigma\rangle\propto t^{-2}$ behaviour (up to relativistic corrections).
    \textit{Bottom}: Comparisons: (i) between the semi-analytic and the MC results (blue); (ii) between the analytic ($f_{\gamma\rm, an.}$) and the MC results (green); (iii) between the MC results with or without additional stable isotopes in the initial ejecta besides ${}^{56}$Ni, as specified in \citet{blondin_standart_2022} (black), showing $<$1\% deviation at relevant times, and larger sensitivity only at the optically thin limit, as $\langle\kappa_{\gamma,E}\rangle$ is sensitive to the composition due to PE absorption of X-rays. The MC computation was done using $\gamma$-ray and X-ray lines from ENDF/B-VIII.0 \citep{brown_endfb-viii0_2018}, and PE and PP opacities from XCOM \citep{berger_xcom_2010} (i.e. not the data provided with URILIGHT, used in Fig. \ref{fig:Semi_vs_MonteCarlo_IA}).}
    \label{fig:typeIa_dep}
\end{figure}

Fig. \ref{fig:typeIa_dep} shows $f_\gamma$ for a Ia SN toy model with pure initial ${}^{56}$Ni ejecta. 
The density profile is taken as that of toy06 \citep{blondin_standart_2022}, which has a column density typical for Ia SN models (Table \ref{table:interpolation}).
$f_\gamma$ computed using a MC simulation \citep[URILIGHT,][with minor modifications, see Appendix \ref{ap:URILIGHT}]{wygoda_type_2019,blondin_standart_2022} agrees to better than 10\% accuracy with the semi-analytic approximation, eqs.~(\ref{eq:f_gamma_all_energies}), (\ref{eq:gamma_E_t_semi})-(\ref{eq:t_E_av}). The maximal underestimation of the semi-analytic approximation at relevant times (before charged particles dominate the heating), occurs near $t_\gamma$, as a result of neglecting multiple Compton scattering\footnotemark.
\footnotetext{At $t\gg t_\gamma$ another underestimation occurs, as the semi-analytic approximation does not include the relativistic effect of "delayed" deposition of $\gamma$-rays accumulated in the ejecta, see late times at Fig. \ref{fig:typeIa_dep} and also at Fig. \ref{fig:Semi_vs_MonteCarlo_KN}. In contrast, for times near $t_\gamma$ the inclusion of relativistic effects in the MC calculation actually reduce the discrepancy between the results, at a 1\% level.}
The analytic approximation $f_{\gamma\rm, an.}(t)$, given by eq. \eqref{eq:f_geff_general} with $t_\gamma$ determined using $\kappa_\geff$ (eq. \ref{eq:t_geff_def}) and $n=2.26$ (Table \ref{table:interpolation}, calculated using eq. \ref{eq:n_interpolation}), is also in better than 10\% agreement with the MC calculation, even well after charged particles begin to dominate the deposition.
$f_\gamma$ deviates from the analytic approximation only at much later times, as it approaches the fraction of energy emitted in X-rays, which are fully deposited until very late times. When the X-rays from ${}^{56}$Co finally lose thermalization efficiency, $f_\gamma$ coincide with the optically thin limit $\langle\kappa_{\gamma,E}\rangle\langle\Sigma\rangle$ (up to relativistic corrections).
It is clear that up to that time, using $\langle\kappa_{\gamma,E}\rangle$ as an effective opacity significantly overestimates the energy deposition. Finally, the figure shows that the inaccuracy introduced by the simplified atomic composition used is negligible, demonstrating the expected validity of the semi-analytic solution with components-averaged $\Bar{\kappa}_{\gamma,E}$ (eqs.~\ref{eq:gamma_E_t_semi}-\ref{eq:t_E_av}) for Ia SNe.

\begin{figure}
    \centering
    \includegraphics[width=\columnwidth]{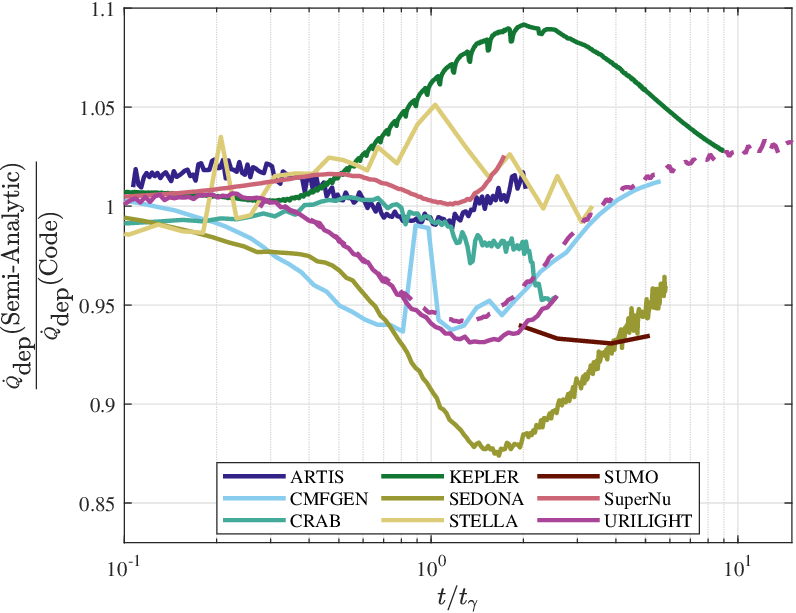}
    \caption{A comparison of the total energy deposition rate $\Dot{Q}_{\text{dep}}(t)$ obtained using our semi-analytic approximation, eqs.~(\ref{eq:f_gamma_all_energies}), (\ref{eq:gamma_E_t_semi})-(\ref{eq:t_E_av}), with the results of simulations using different transport codes, from \citet{blondin_standart_2022}, for the Ia SN toy model - toy06, with $t_\gamma \approx 38.8$ d (similar results were obtained for toy01).
    Solid lines were computed directly using $\Dot{Q}_{\text{dep}}(t)$ published by \citet{blondin_standart_2022}. The dashed line is a corrected URILIGHT result, following minor modifications to the code (see Appendix \ref{ap:URILIGHT}).
    The simulation results were computed for the detailed atomic composition of toy06, while the semi-analytic approximation assumes an initially pure ${}^{56}$Ni composition.
    The $\sim$3\% deviation of URILIGHT at late times is due to $\sim$3\% overestimation of the positron energy release rate compared to ENDF data \citep{brown_endfb-viii0_2018}.
    }
     \label{fig:Semi_vs_MonteCarlo_IA}
\end{figure}

In Fig. \ref{fig:Semi_vs_MonteCarlo_IA} we show the deviations between $\Dot{Q}_{\text{dep}}$ computed using the semi-analytic approximation, eqs.~(\ref{eq:f_gamma_all_energies}), (\ref{eq:gamma_E_t_semi})-(\ref{eq:t_E_av}), and the results of simulations presented by \citet{blondin_standart_2022}, which were performed for the same Ia SN toy model - toy06 (similar results were obtained for toy01), and using different $\gamma$-ray transport codes: MC codes (ARTIS, CMFGEN, SEDONA, URILIGHT) and gray transport codes (CRAB, KEPLER, SUMO, STELLA, SuperNu).
The semi-analytic approximation agrees with the results of most codes to a few percent accuracy, showing in particular that the level of inaccuracy introduced by the approximation is even slightly smaller than the $\sim$10\% variation between MC codes (that exists in particular between ARTIS and SEDONA).
For KNe, the $<$10\% inaccuracy introduced by our approximations is well within the uncertainties due to nuclear physics uncertainties and due to numeric transport calculations' inaccuracies.

\section{The $\gamma$-ray deposition in Kilonovae}
\label{sec:KNe}

In this section, we study the $\gamma$-ray deposition in the radioactive matter ejected from NSM. First, we examine the characteristic $\kappa_\geff$, that determines $t_\gamma$. To this end, we consider ejecta with uniform atomic/isotopic composition ($\rho_{\rm rad}=\rho$), with a wide range of $\{Y_e,s_0,\rho t^3\}$ values.
In \S~\ref{subsec:KNe_model} we describe the nucleosynthesis and radioactive decay calculations. In \S~\ref{subsec:KNe_kappa_geff} we show the results for $\kappa_\geff$, calculated using eq. \eqref{eq:k_geff_integral_def_2} with $\phi_\gamma(E,t)$ and $\kappa_{\gamma,E}$ obtained from the calculations of the nucleosynthesis and radioactive decay, and in \S~\ref{subsec:nuclear_sensitivity} we show that $\kappa_\geff$ is not sensitive to nuclear physics uncertainties. Then, in \S~\ref{subsec:comp_numerical} we demonstrate the accuracy of our approximations for $f_\gamma$ by comparing the analytic and semi-analytic results to the results of MC calculations. Finally, in \S~\ref{sec:comp_others} we compare our results to those of earlier studies.

\subsection{Nucleosynthesis and radioactive decay calculations}
\label{subsec:KNe_model}
The atomic and isotopic composition of the ejecta is determined by a nucleosynthesis computation, performed using the publicly available code, \textit{SkyNet} \citep{lippuner_skynet_2017}.
It is assumed that the ejecta is initially in a nuclear statistical equilibrium (NSE) state, at which the composition is determined by nuclear masses, the thermodynamic conditions and $Y_e$ \citep[e.g.][]{meyer_entropy_1993,cowan_origin_2021}.
As NSE requires high temperatures of $\gtrsim$ few GK, we set the initial temperature as $T_0 = 10$ GK.
Such high temperatures, that allow for NSE, are expected following NSM, as even if the matter is not heated hydrodynamically, nuclear reactions on timescales much shorter than the expansion/r-process timescales will deposit $\sim$1 MeV per baryon into the plasma and heat it up to NSE temperatures \citep{freiburghaus_r-process_1999}. The expansion of the ejecta evolves with time from an expansion at which $\rho$ evolves slower than $t^{-3}$ soon after the merger (the result of high temperatures causing pressure waves to be faster than material velocities), to an homologous expansion at which $\rho \propto t^{-3}$ (the result of ejecta cooling leading to supersonic material velocities).
This evolution is approximated as \citep{lippuner_r-process_2015}
\begin{equation}
    \rho(t) = 
    \begin{cases}
    \rho_0 e^{-t/\tau} & \text{for } t\leq 3 \tau, \\
    \rho_0 \Big( \frac{3 \tau}{et} \Big)^3              & \text{otherwise,}
\end{cases}
 \label{eq:skynet_density_history}
\end{equation}
\noindent where $\rho_0$ is the initial density and $\tau$ is the expansion timescale, or the timescale for onset of homologous expansion.
The ejecta is parameterized by the asymptotic (homologous phase) $\rho t^3$ and by uniform $Y_e$ and $s_0$.
With this parameterization, and with the fixed $T_0$, we determine $\rho_0$ by the equation of state of the plasma, and set $\tau$ to agree with $\rho t^3$ asymptotically, according to eq. \eqref{eq:skynet_density_history}\footnote{Note that in our approach $\tau$ is not a physically meaningful quantity, as it is determined by $T_0$ through $\rho_0$.
Some earlier works which used \textit{SkyNet} to survey the NSM parameters space \citep{lippuner_r-process_2015,perego_production_2022} used $\tau$ as a physical parameter, and thus the interpretation of their results is complicated by a non-trivial and undesirable dependence on $T_0$. Our approach is the same as that employed by works that used $\tau$, but determined it from the asymptotic $\rho t^3$ values obtained for hydrodynamic trajectories from NSM simulations \citep[e.g.][]{radice_dynamical_2016,radice_binary_2018,nedora_numerical_2021,perego_production_2022}.}.

\textit{SkyNet} includes 7843 isotopes, up to $^{337}\textrm{Cn}$. We use the same nuclear physics input methodology described by \citet{lippuner_skynet_2017}. Specifically, we employ the latest version (and not the version distributed with \textit{SkyNet}) of the JINA REACLIB database \citep{cyburt_jina_2010}, with few corrections of incorrect half-lives described in Appendix \ref{ap:REACLIB}.
Strong inverse rates are computed assuming detailed balance. We use the \texttt{sym0} version of the neutron-induced fission rates, and the spontaneous fission rates taken from \citet{roberts_electromagnetic_2011}, both distributed with \textit{SkyNet}.
Nuclear masses are taken from the REACLIB database, that includes experimental values and theoretical masses based on the FRDM model \citep{moller_nuclear_2016}.

Radioactive decay data (half-lives and decay products) are taken from the latest version of the experiments-based database, ENDF/B-VIII.0 \citep{brown_endfb-viii0_2018}.
We use the $\gamma$-rays (and X-rays) line spectra provided\footnote{We ignore isotopes which have continuous $\gamma$-ray data in the database. Their contribution to the energy released is negligible.} to calculate the $\gamma$-ray (and X-ray) injection rate $d\Dot{N}_{\gamma}/dE$.
We also compute $\dot{Q}_{\text{charged}}$ in order to compute the total energy deposition rate $\Dot{Q}_{\text{dep}}$.
To do so we use \textit{BetaShape} \citep{mougeot_betashape_2017} to calculate $e^{\pm}$ $\beta$-decays' spectra and use the line spectrum of alphas given by ENDF. 
To account for energy carried by SF fragments, we also include the Q-value of SF (assuming all of it is carried by the fission fragments).

We note the following two points, which are further discussed in Appendix \ref{ap:Checks}, regarding the validity of our calculations. First, even tough the early ($\sim$1 s) composition of the ejecta contain very neutron-rich isotopes which lack experimental radioactive data, by about 1 hour after the merger the ejecta is dominated by isotopes which are close to the valley of stability and have experimental decay data. Since we consider times later than 1 hour, our radioactive decay energy release and $\gamma$-ray spectra calculations are not subject to large uncertainties due to the computed composition. 

Second, the SF rates supplied by \textit{SkyNet} are erroneously large compared to the true values, as they are based on the total nuclide activity, without taking into account the (usually very small) branching ratio to SF.
As a result, the energy release rate given as output by \textit{SkyNet} (which we do not use) overestimates the true energy release rate at low $Y_e$ and $t\gtrsim10$ d, as SF energy release rate wrongly dominates the total energy release \citep[this dominance was disscused by][in their section 2.6]{lippuner_r-process_2015}.
We calculate the energy release rate from the resulting composition using the correct branching ratios, making sure that the effect of the wrong SF rates on the results is negligible.

\subsection{The effective $\gamma$-ray opacity of Kilonovae}
\label{subsec:KNe_kappa_geff}

In this subsection, we explore the dependence of $\kappa_\geff$, as defined in eq. \eqref{eq:k_geff_integral_def_2}, on the parameters defining the KN ejecta. $\kappa_\geff$ depends on the $\gamma$-ray spectrum, $\phi_\gamma$, and opacities, $\kappa_{\gamma,E}$, and on the average column density, $\langle\Sigma\rangle$. $\phi_\gamma$ and $\kappa_{\gamma,E}$ depend on the composition, which is determined by $\{Y_e,s_0,\rho t^3\}$. We explore a wide range of values of these parameters, $0.05\le Y_e\le0.45$, $1\le s_0[\kb]\le100$, and $10^{-3}\le [\rho t^3]_{\text{KN}} \le10^3$ (see eq. \ref{eq:rhoKN}). To explore a wide range of column densities for a given density, we parameterize $\langle\Sigma\rangle$ as $\langle\Sigma\rangle t^2=\rho t^3\Delta \text{v}$,
and explore $0.01<\Delta \text{v}/\text{c}<1$.
For each set of ejecta parameter values, $\kappa_\geff$ is obtained by solving eq. \eqref{eq:k_geff_integral_def_2}, with $\phi_{\gamma}(E,t)$ and $\kappa_{\gamma,E}$ obtained from the numeric nucleosynthesis calculation described in \S~\ref{subsec:KNe_model}, and $\langle\Sigma\rangle t^2 = \rho t^3 \Delta \text{v}$.

We find $\kappa_\geff$ to be close to the effective Compton scattering opacity $\approx$0.025\cmgr in all cases, with a weak dependence on $Y_e$,
\begin{equation}
    \kappa_\geff\approx \begin{cases}
    0.05 \text{\cmgr}, & \text{$Y_e \lesssim Y_{e,\text{th}}$},\\
    0.03 \text{\cmgr}, & \text{$Y_e \gtrsim Y_{e,\text{th}}$}.
  \end{cases}
  \label{eq:kilonova_kappa_approx}
\end{equation}
\noindent The threshold electron fraction, $Y_{e,\text{th}}$, varies slightly with $s_0$,
\begin{equation}
    Y_{e,\text{th}} (s_0) \approx \begin{cases}
    0.20, & \text{$s0\lesssim30 \kb$},\\
    0.25, & \text{$s0\gtrsim40 \kb$},
  \end{cases}
  \label{eq:kilonova_Ye_th}
\end{equation}
\noindent and it coincides with the threshold required for strong r-process nucleosynthesis, at which nuclei beyond the second r-process peak are produced. These include lanthanides, third r-process peak isotopes and actinides \citep[e.g.][]{perego_r-process_2021}.
The time $t_\gamma$, defined in eq. \eqref{eq:t_geff_def} using $\kappa_\geff$, is thus given by
\begin{equation}
 t_\gamma\approx [\langle\Sigma\rangle t^2]_{\text{KN}}^{\frac{1}{2}} \begin{cases}
    1.2\text{ d}, & \text{$Y_e \lesssim Y_{e,\text{th}}$},\\
    0.95\text{ d}, & \text{$Y_e \gtrsim Y_{e,\text{th}}$}.
  \end{cases}
  \label{eq:kilonova_t_geff}
\end{equation}
Here,
\begin{equation}
    \label{eq:SigKN}
    [\langle\Sigma\rangle t^2]_{\text{KN}} \equiv [\rho t^3]_{\text{KN}}\frac{\Delta \text{v}}{0.2\text{c}} =  \frac{\langle\Sigma\rangle t^2}{2.2\times 10^{11} {\text{g}}\ {\text{cm}}^{-2}\ \text{s}^{2}}
\end{equation}
is the column density normalized to values inferred for the ejecta associated with GW170817, $[\rho t^3]_{\text{KN}} = 1$, $\Delta \text{v} = 0.2 \text{c}$.

The results summarized in eqs. (\ref{eq:kilonova_kappa_approx}, \ref{eq:kilonova_Ye_th}, \ref{eq:kilonova_t_geff}) are illustrated in the following figures.
Fig. \ref{fig:kilonova_kappa_eff_rhot3=1} shows $\kappa_\geff$ as a function of $s_0$ and $Y_e$, for $[\rho t^3]_{\text{KN}}=1$  and $\Delta \text{v}=0.2\text{c}$. In the map of $\kappa_\geff$ as a function of $Y_e$ and $s_0$, showing regions separated by $Y_{e,\text{th}} (s_0)$, it is clear that the variation in each region is small compared to the variation between the regions.
Similar results are shown in Figs. \ref{fig:kilonova_kappa_all_dependence:a} and \ref{fig:kilonova_kappa_all_dependence:b}, which show maps of $\kappa_\geff$ as a function of $Y_e$ and $s_0$ for two additional values of $\rho t^3$: $[\rho t^3]_{\text{KN}} = 0.1$ and $[\rho t^3]_{\text{KN}} = 10$. 

The weak dependence of $\kappa_\geff$ on $\langle\Sigma\rangle t^2 = \rho t^3\Delta \text{v}$ is demonstrated in Figs. \ref{fig:kilonova_kappa_all_dependence:a} and \ref{fig:kilonova_kappa_all_dependence:b}, and further in Figs. \ref{fig:kilonova_kappa_all_dependence:c} and \ref{fig:kilonova_kappa_all_dependence:d}, which show $\kappa_\geff$ as a function of $[\rho t^3]_{\text{KN}}$ for fixed values of $\Delta \text{v}$ and $s_0$, and as a function of $\Delta \text{v}$ for $[\rho t^3]_{\text{KN}}=1$ and fixed $s_0$, respectively (Note that changing $\rho t^3$ is accompanied by a change in composition, while changing $\Delta \text{v}$ changes only $\langle\Sigma\rangle t^2$). Clearly, the effect of changing either of these causes only a weak variation in $\kappa_\geff$.

Fig. \ref{fig:kilonova_t_gamma_groups} shows that for all ejecta parameter values considered, $t_\gamma$ agrees with eq. \eqref{eq:kilonova_t_geff} (there is a small overestimation at low $Y_e$ and $[\rho t^3]_{\text{KN}}\leq0.1$, as there $\kappa_\geff\approx0.04$\cmgr). Fig. \ref{fig:kilonova_kappa_gamma_mean} compares $\kappa_\geff$ to the energy-weighted mean opacity $\langle\kappa_{\gamma,E}\rangle$ (eq. \ref{eq:f_gamma_late}). As discussed in \S~\ref{sec:intro}, $\langle\kappa_{\gamma,E}\rangle$ significantly overestimates $\kappa_\geff$, due to incorrect weighting of low-energy (and high-opacity) $\gamma$-rays (and X-rays), especially in high-$Z$ (low-$Y_e$) ejecta at which low-energy $\gamma$-rays have large $\kappa_{\gamma,E}$. Notably, discarding the X-rays from the calculation of $\langle\kappa_{\gamma,E}\rangle$ (that resolved the overestimation for Ia SNe, see \S~\ref{sec:type_ia}), still results in a large overestimation of $\kappa_\geff$ as even low-energy $\gamma$-rays (not X-rays) have large $\kappa_{\gamma,E}$ at high-$Z$ material. However, as in Ia SNe, the overestimation is reduced to some extent by using $\langle\kappa^{1/2}_{\gamma,E}\rangle^2$, without the need to discard the X-rays.

\begin{figure}
\centering
 \includegraphics[width=\columnwidth]{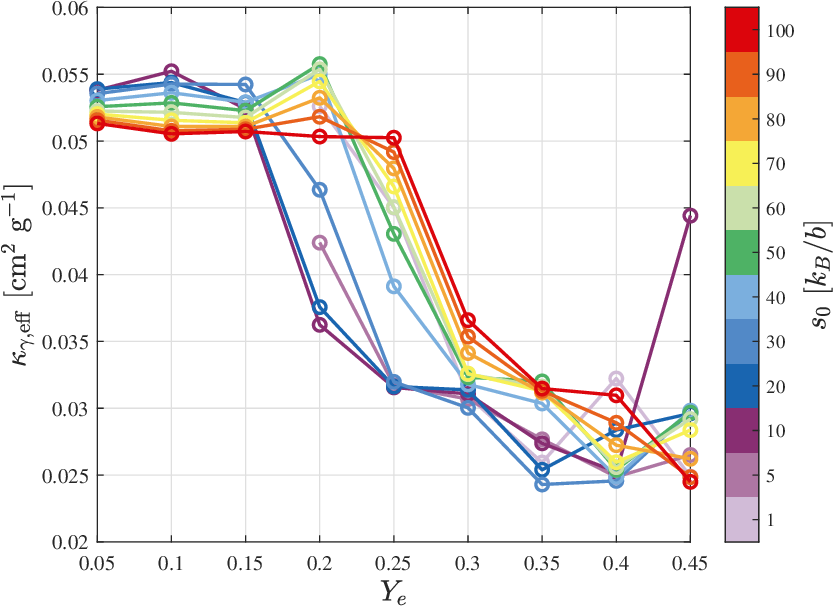}
\centering
 \includegraphics[width=\columnwidth]{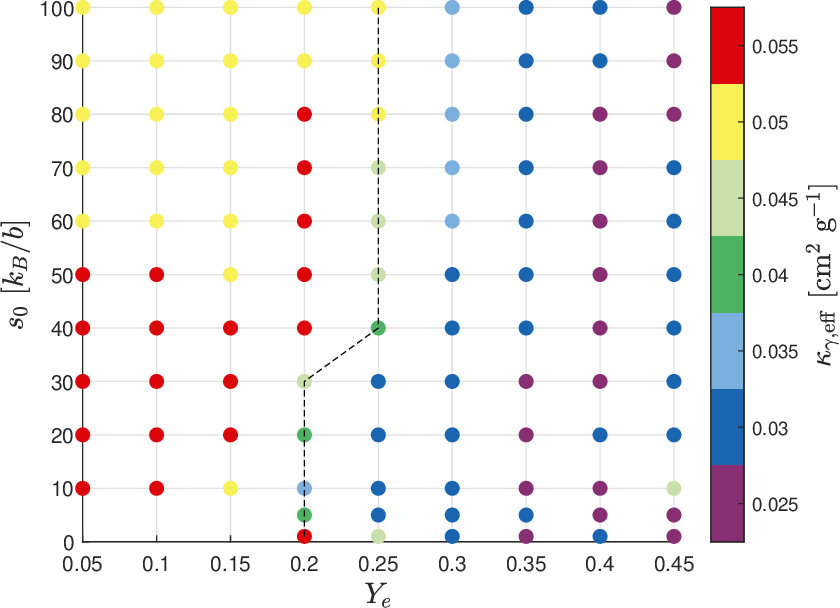}
 \caption{$\kappa_\geff$ for KNe with $[\rho t^3]_{\text{KN}} = 1$ (eq.~\ref{eq:rhoKN}) and $\Delta \text{v} = 0.2 \text{c}$, as a function of $Y_e$ and $s_0$. \textit{Top:} $\kappa_\geff$ curves of constant $s_0$ (in color) as a function of $Y_e$. \textit{Bottom:} a map of $\kappa_\geff$ as a function of $Y_e$ and $s_0$. The dashed line shows the threshold curve $Y_{e,\text{th}}(s_0)$, given by eq. \eqref{eq:kilonova_Ye_th}. Missing points are due to failed runs of \textit{SkyNet}.}
 \label{fig:kilonova_kappa_eff_rhot3=1}
\end{figure}
\begin{figure*}
    \centering
    \begin{subfigure}[t]{0.45\textwidth}
        \centering
        \includegraphics[width=\linewidth]{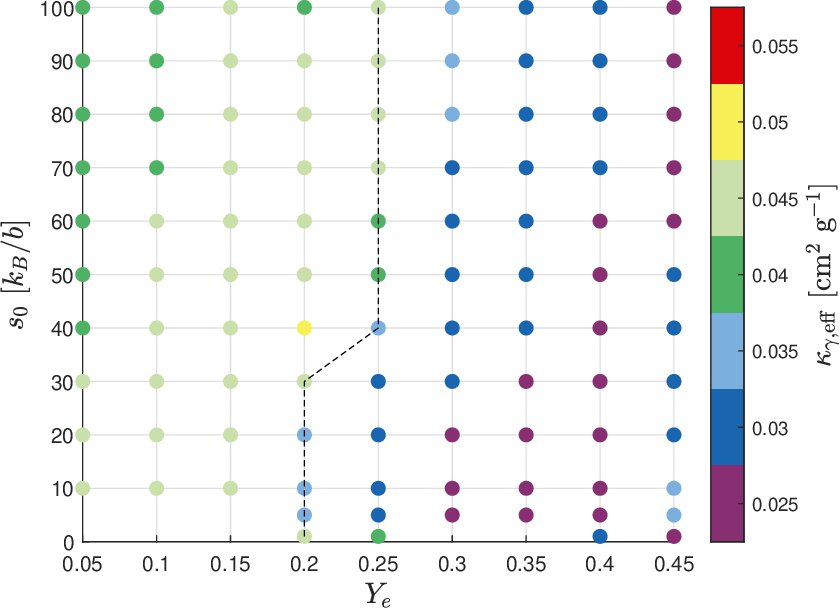} 
        \caption{$[\rho t^3]_{\text{KN}} = 0.1$}
        \label{fig:kilonova_kappa_all_dependence:a}
    \end{subfigure}
    \hfill
    \begin{subfigure}[t]{0.45\textwidth}
        \centering
        \includegraphics[width=\linewidth]{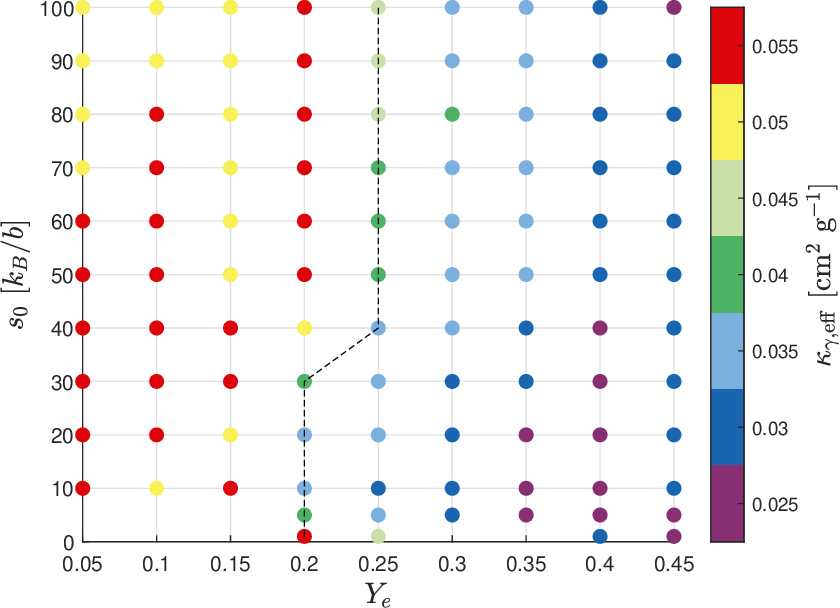} 
        \caption{$[\rho t^3]_{\text{KN}} = 10$}
        \label{fig:kilonova_kappa_all_dependence:b}
    \end{subfigure}
    \centering
    \begin{subfigure}[t]{0.45\textwidth}
        \centering
        \includegraphics[width=\linewidth]{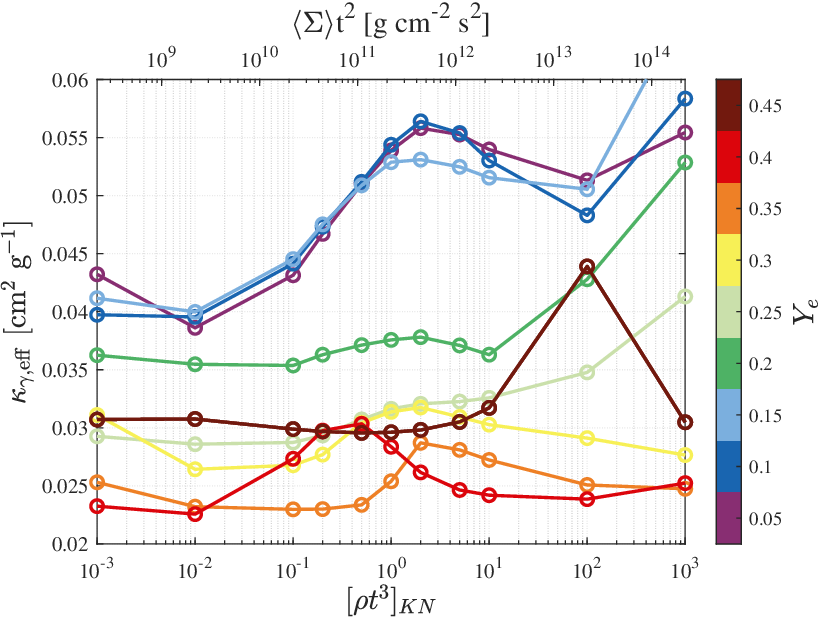} 
        \caption{$s_0 = 20~\kb$ and $\Delta \text{v} = 0.2 \text{c}$}
        \label{fig:kilonova_kappa_all_dependence:c}
    \end{subfigure}
    \hfill
    \begin{subfigure}[t]{0.45\textwidth}
        \centering
        \includegraphics[width=\linewidth]{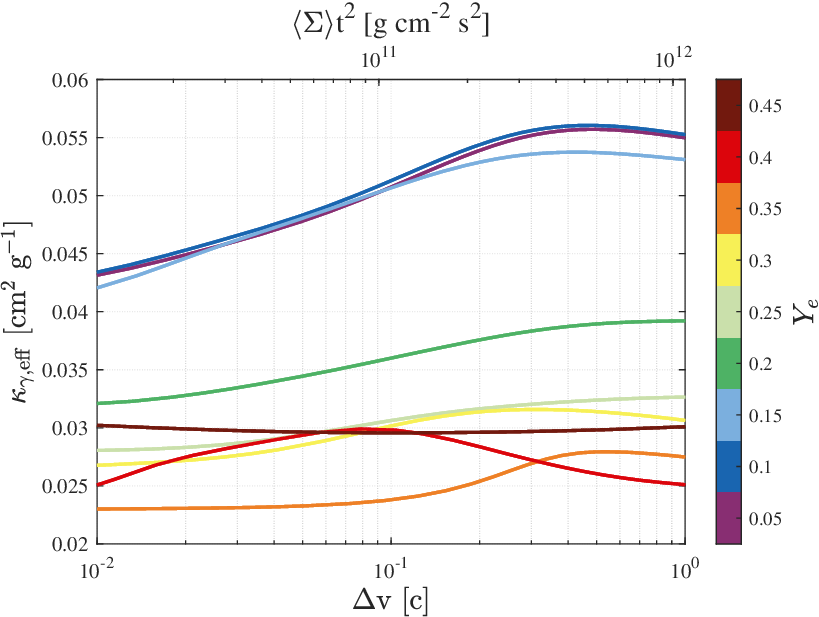} 
        \caption{$s_0 = 20~\kb$ and $[\rho t^3]_{\text{KN}} = 1$}
        \label{fig:kilonova_kappa_all_dependence:d}
    \end{subfigure}
\caption{The dependence of $\kappa_\geff$ on ejecta parameters.
Panels (a) and (b): maps of $\kappa_\geff$ as a function of $Y_e$ and $s_0$ for ejecta with  $\Delta \text{v} = 0.2 \text{c}$ and $[\rho t^3]_{\text{KN}} = 0.1$ and 10 (eq.~\ref{eq:rhoKN}). $\kappa_\geff$ exhibits the same threshold behaviour near $Y_{e,\text{th}}(s_0)$ (dashed line) as shown in Fig.~\ref{fig:kilonova_kappa_eff_rhot3=1} for $[\rho t^3]_{\text{KN}} = 1$. Panel (c) shows the $\rho t^3$ dependence of $\kappa_\geff$, for $s_0=20~\kb$, $\Delta \text{v}=0.2\text{c}$ and different $Y_e$ values. Panel (d) shows the $\Delta \text{v}$ dependence of $\kappa_\geff$ (through the change in $\langle\Sigma\rangle t^2 = \rho t^3 \Delta \text{v}$), for $[\rho t^3]_{\text{KN}} = 1$, $s_0=20~\kb$ and different $Y_e$ values. The upper axes in panels (c) and (d) show the column density $\langle\Sigma\rangle t^2 = \rho t^3 \Delta \text{v}$.
The overshooting blue point in the right upper corner of panel (c) is $\kappa_\geff \approx 0.067$\cmgr.}
 \label{fig:kilonova_kappa_all_dependence}
\end{figure*}

\begin{figure}
\centering
 \includegraphics[width=\columnwidth]{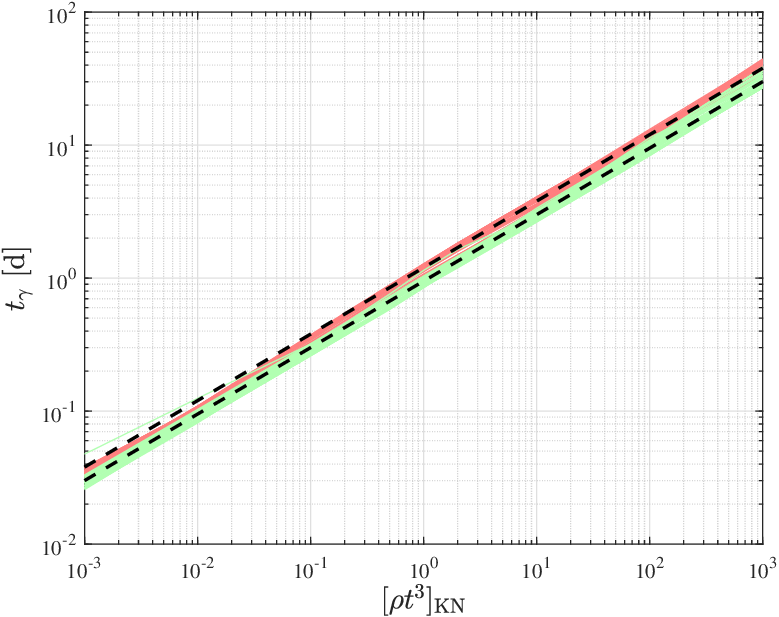}
 \caption{$t_\gamma$, given by eq.~(\ref{eq:t_geff_integral_def_2}) or, equivalently, eq.~(\ref{eq:t_geff_def}), as a function of $[\rho t^3]_{\text{KN}}$ (eq.~\ref{eq:rhoKN}) for all $\{Y_e,s_0,\rho t^3\}$ values and $\Delta \text{v} = 0.2 \text{c}$. Results for neutron-poor conditions, $Y_e>Y_{e,\text{th}} (s_0)$ are shown in green, results for neutron-rich conditions, $Y_e\leq Y_{e,\text{th}} (s_0)$ are shown in red. The dashed lines show the analytic approximation given by equation \eqref{eq:kilonova_t_geff}.
 The overshooting green line at low $[\rho t^3]_{\text{KN}}$ is of $Y_e = 0.45$ and $s_0 = 5~\kb$.}
 \label{fig:kilonova_t_gamma_groups}
\end{figure}

\begin{figure}
\centering
 \includegraphics[width=\columnwidth]{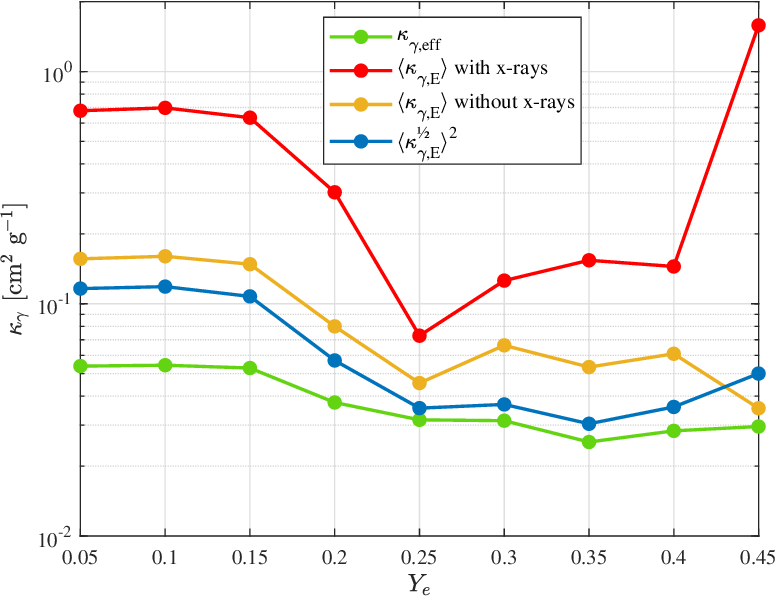} 
 \caption{The results of different definitions of an effective $\gamma$-ray opacity, for KNe with $[\rho t^3]_{\text{KN}}=1$ (eq.~\ref{eq:rhoKN}), $\Delta \text{v} = 0.2 \text{c}$, $s_0=20~\kb$ and different $Y_e$ values. The green line shows $\kappa_\geff$ as defined by eq. \eqref{eq:k_geff_integral_def_2}; the red line shows the energy-weighted average $\langle\kappa_{\gamma,E}\rangle$, including both $\gamma$-rays and X-rays (computed at the time at which 
 $\langle\kappa_{\gamma,E}\rangle(t)\langle\Sigma\rangle t^2 = 1$);
 the gold line shows the energy-weighted average $\langle\kappa_{\gamma,E}\rangle$, discarding the X-rays; the blue line shows $\langle\kappa^{1/2}_{\gamma,E}\rangle^2$ (corresponded to $t_\gamma = \langle t_{\gamma,E}\rangle$; computed at $t$ for which $\langle\kappa^{1/2}_{\gamma,E}\rangle^2(t)\langle\Sigma\rangle t^2 = 1$).
 Clearly, $\langle\kappa_{\gamma,E}\rangle$ overestimates $\kappa_\geff$, and hence $t_\gamma$ and $f_\gamma$.
 }
 \label{fig:kilonova_kappa_gamma_mean}
\end{figure}

Fig. \ref{fig:kilonova_kappa_eff_examination} shows $\kappa_\geff$ as a function of $Y_e$ together with the effective atomic number $Z_{\rm eff}$ of the ejecta\footnotemark and the energy-weighted mean energy of $\gamma$-rays,
both evaluated at 1 d (approximately $t_\gamma$). 
\footnotetext{Motivated by the rough scaling $\kappa_\text{PE}$$\sim$$Z^{2.5}$, $Z_{\rm eff}$ is defined as
\begin{equation}
    Z_{\rm eff} = [\Sigma_{Z} X_Z Z^{2.5}]^{1/2.5},
    \label{eq:Z_eff}
\end{equation}
where $X_Z$ is the mass fraction of element $Z$. The $\kappa_{\gamma,E}$ of the ejecta is well approximated (at a few percent accuracy) by the opacities of the element $Z\approx Z_{\rm eff}$ (with its natural isotopic abundance, as at relevant times the nuclei are close to the valley of stability).}
$Z_{\rm eff}$ increases as $Y_e$ is decreased below $Y_{e,\text{th}}$, due to strong r-process contribution,
leading to opacity increase due to increased PE contribution ($\kappa_{\rm PE}$$\sim$$Z^{2.5}$). 
Examining the mean $\gamma$-ray energy dependence on $Y_e$ shows that while the spectrum is softened as $Y_e$ is decreased (discarding the outlier at the very neutron-poor case of $Y_e = 0.45$), as is expected since higher-$Z$ nuclei tend to emit softer $\gamma$-rays, the mean energy is near $\sim$1 MeV for all $Y_e$ values. This implies that $\kappa_\geff$ is only weakly affected by the change in $\gamma$-ray spectrum, and that the increase of $\kappa_\geff$ at low $Y_e$ is due mainly to the increase in $Z$.
Note, that both $Z_{\rm eff}$ and the mean $\gamma$-ray energy are saturated at low $Y_e$. This is due to fission-cycling, which leads to a robust (nearly initial conditions independent) abundance pattern \citep[e.g.][]{perego_r-process_2021}.

The (weak) entropy dependence of $Y_{e,\text{th}}$ in eq. \eqref{eq:kilonova_Ye_th} is due to the fact that for a given $Y_e$, higher $s_0$ leads to NSE composition of lighter nuclei \citep{meyer_entropy_1993}. This increases the abundance of neutrons while reducing the abundance of seed nuclei (the nuclei that capture neutrons in the r-process), thus allowing for heavy elements formation (strong r-process) at higher $Y_e$ \citep[see][for reviews]{cowan_origin_2021,perego_r-process_2021}.
In contrast, at very low initial entropy $s_0$$\lesssim$1$\kb$, $Y_{e,\text{th}}$ is larger (Fig. \ref{fig:kilonova_kappa_eff_rhot3=1}), as the seed abundance is heavy enough to allow the production of some nuclei beyond the second r-process peak \citep{lippuner_r-process_2015}.

\begin{figure}
\centering
 \includegraphics[width=\columnwidth]{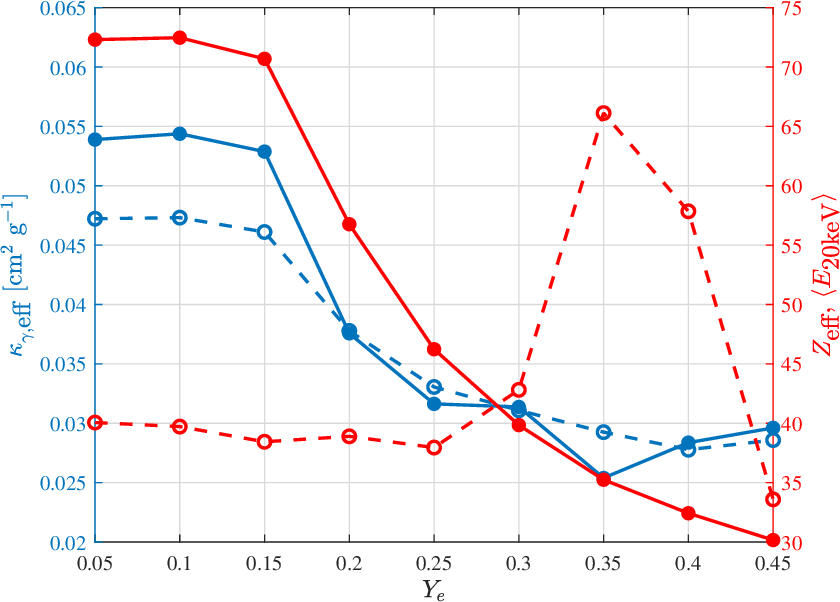}
 \caption{The $Y_e$-dependence of $\kappa_\geff$ (solid blue line), for KNe with $[\rho t^3]_{\text{KN}}=1$ (eq.~\ref{eq:rhoKN}), $\Delta \text{v} = 0.2 \text{c}$ and $s_0=20~\kb$, shown together with the effective atomic number of the ejecta, $Z_{\rm eff}$ (eq.~\ref{eq:Z_eff}) (solid red line), and the energy-weighted average energy of $\gamma$-rays, $\langle E \rangle = \int{dE E\phi_\gamma(E,t)}$, in units of 20keV (dashed red line) (both evaluated at 1 d, approximately $t_\gamma$). In dashed blue, we show $\kappa_\geff$ calculated for $\phi_\gamma(E,t)$ fixed (at all times and $Y_e$ values) to be $\phi_\gamma(E,t=1\text{ d})$ for $Y_e = 0.4$. The graphs demonstrate that the dependence of $\kappa_\geff$ on $Y_e$ is due mainly to the dependence of $Z_{\rm eff}$ (rather than of $\phi_\gamma(E,t)$) on $Y_e$.
 }
 \label{fig:kilonova_kappa_eff_examination}
\end{figure}

The highest $Y_e$ case, $Y_e=0.45$, is anomalous (see e.g. Figs. \ref{fig:kilonova_kappa_eff_rhot3=1}, \ref{fig:kilonova_kappa_gamma_mean}), in the sense that the neutron-to-seed ratio is above unity only at the higher end of our $s_0$ range, $s_0\gtrsim100\kb$, and thus no r-process occurs at lower $s_0$ \citep{farouqi_charged-particle_2010,cowan_origin_2021}.

\subsection{Sensitivity to nuclear physics uncertainties}
\label{subsec:nuclear_sensitivity}
The rapid neutron capture process which occurs at neutron-rich conditions following NSM creates neutron-rich nuclei, far from the valley of stability, all the way to the neutron drip line.
Basic properties of most of these nuclei were not measured, such as the nucleus mass, its $\beta$-decay rate and its neutron capture cross section \citep[e.g.][]{cowan_origin_2021}. Consequently, theoretical nuclear physics models are used to determine the nuclear masses and reaction rates when empirical values are unavailable.
For example, \textit{SkyNet} uses the REACLIB nuclear database \citep{cyburt_jina_2010}, which includes nuclear masses and interaction rates calculated using the FRDM mass model.
Directly, the masses of the nuclei enter the calculation in determining the abundance in NSE \citep[through a Boltzmann factor with the nuclear binding energy, e.g.][]{meyer_entropy_1993} and inverse reaction rates (through detailed balance), while the interaction rates are used to evolve the abundance outside of NSE.
In this subsection, we examine the sensitivity of $\kappa_\geff$ to these nuclear physics uncertainties.

In Fig. \ref{fig:kilonova_random_rates_and_mass_model} we show the variation of $\kappa_\geff$ as we modify the theoretical nuclear rates used in \textit{SkyNet}. Specifically, this was done by using 100 generated realizations of the REACLIB nuclear reaction rates, in which every strong or weak interaction rate that is labeled as a theoretical rate\footnote{We included mo97, that is wrongly labeled by REACLIB as experimental.} was changed by a factor $C$, $\lambda\rightarrow C\lambda$, where $C$ is log-uniformly distributed over $[10^{-2}, 10^2]$ \citep[as simlarly done by past works, e.g.][]{surman_neutron_2009,surman_sensitivity_2014,mumpower_sensitivity_2014}.
Further, to test the direct sensitivity to the nuclear mass model, the same rates realizations were run using the UNEDF1 mass model \citep[][\url{http://massexplorer.frib.msu.edu}]{erler_limits_2012,kortelainen_nuclear_2012}, instead of the usual FRDM nuclear mass model of \textit{SkyNet}.
Relative to the "nominal" value of $\kappa_\geff$, the standard deviation of the change in $\kappa_\geff$ is $\lesssim$10\%.
Only in a single realization (at $Y_e = 0.25$, $[\rho t^3]_{\text{KN}} = 10$ and UNEDF1 mass model) a much larger variation was observed (120\%).

The weak sensitivity of $\kappa_\geff$ to  orders of magnitude variations in the nuclear interaction rates is the result of the fact that as long as the KN is powered by an ensemble of nuclei, most of the $\gamma$-ray energy is expected to be emitted as $\sim$1~MeV $\gamma$-rays, and thus $\kappa_\geff$ never deviates much from the effective Compton scattering opacity  $\approx 0.025$\cmgr.

While these results show that our approximations for $f_\gamma$ are robust to nuclear physics uncertainties, $\Dot{Q}_{\text{dep}}$ itself is not.
This is due to the much larger uncertainty in the energy production rates, $\Dot{Q}_{\gamma}$ and $\Dot{Q}_{\text{charged}}$. For example, within our sample of calculations, we found a variation which is mostly up to a factor of 2 in $\Dot{Q}_{\gamma}(t=1\text{ d})$, similar to that found in past works \citep[e.g.][figure 3]{zhu_modeling_2021}.
Note that while our analysis examines large nuclear physics variations, it is limited and does not capture the entire scope of these uncertainties\footnotemark.
However, including these extensions will not change our results for $\kappa_\geff$, eqs. (\ref{eq:kilonova_kappa_approx}, \ref{eq:kilonova_Ye_th}), since as noted above, $\sim$1 MeV $\gamma$-rays are still expected to dominate the spectrum.
\footnotetext{The limitations include: (i) Using only 100 realizations, which cannot capture the extent of variation for the few 10,000 theoretical rates of REACLIB; (ii) There may exist theoretical rates that are mislabeled as experimental and are hence not part of our examination (such as bb92, which was not included in this analysis, but most likely refer to theoretical rates);
(iii) Rates of induced or spontaneous fission were not changed \citep[in some cases, e.g. the HFB mass model of][fission can substantially affect the energy production]{zhu_modeling_2021}; (iv) The effect of (long-lived) isomers was not included (these are absent in \textit{SkyNet}); (v) The nuclear mass model and the rates were changed independently, in contrast to the self-consistent analysis done in some earlier works \citep[e.g.][]{mumpower_impact_2015,zhu_modeling_2021}.
}

\begin{figure}
 \centering
 \includegraphics[width=\columnwidth]{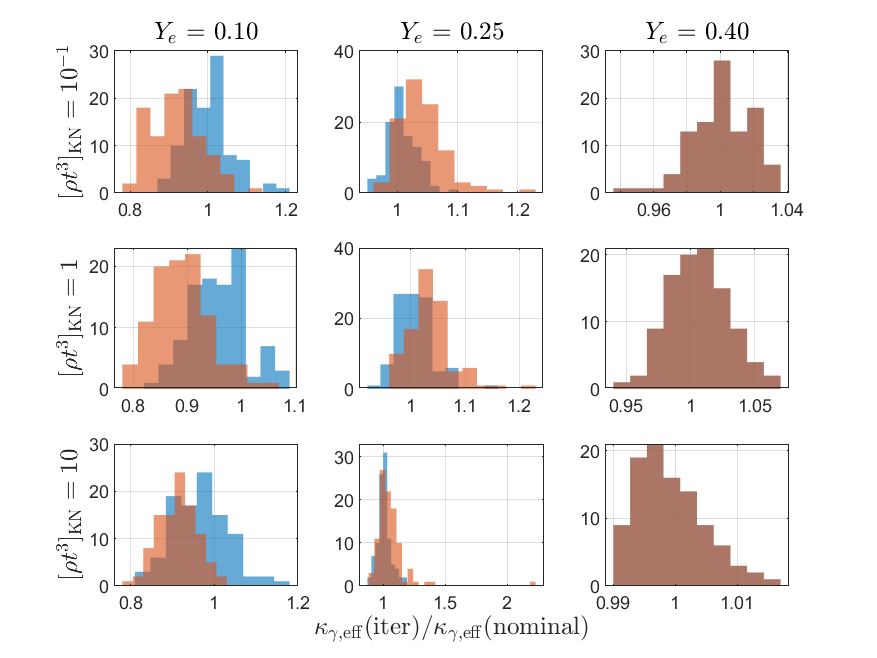}
 \caption{Counts histogram of the ratio between $\kappa_\geff$ obtained for a random modification of the nuclear rates and the nominal value of $\kappa_\geff$. The results are shown for $s_0 = 20 \kb$ and $\Delta \text{v} = 0.2\text{c}$, and for various values of $Y_e$ and $[\rho t^3]_{\text{KN}}$ (eq. \ref{eq:rhoKN}). Two mass models have been used: FRDM (the fiducial mass model, blue) and UNEDF1 (orange).}
 \label{fig:kilonova_random_rates_and_mass_model}
\end{figure}

\subsection{Comparison to detailed numerical calculations}
\label{subsec:comp_numerical}

In this subsection we compare $f_\gamma(t)$ obtained by MC $\gamma$-ray transport calculations to our semi-analytic and analytic approximations for $f_\gamma$.
We consider spherically symmetric ejecta of uniform composition and varying density profiles.
We have examined several types of density profiles: a uniform density shell (with mass extending from $0.1\text{c}$ to $0.3\text{c}$); the density profiles of \citet{barnes_effect_2013}, \citet{barnes_radioactivity_2016}, \citet{barnes_kilonovae_2021} (with a characteristic velocity $\text{v}_\text{char}$ taken as 0.1c or 0.2c); the density profiles of \citet{waxman_constraints_2018}. All yielded similar results. Since the profiles of \citet{waxman_constraints_2018} are characterized by the largest velocity spread, and hence by the largest relativistic effects, they serve as the most stringent test of our approximations. We, therefore, show below the results obtained for these profiles: ejecta of total mass $M$, with minimal velocity $\text{v}_\text{M}$ and a power-law dependence of mass on velocity, $m(\geq\text{v}) \propto \text{v}^{-1/\alpha}$ truncated at $\text{v}=c$.

For each set of $\{Y_e,s_0,\rho t^3\}$ values, we construct an ejecta with the numerically calculated composition (as described in \S~\ref{subsec:KNe_model}), and set\footnotemark $M=0.05 [\rho t^3]_{\text{KN}} M_\odot$, $\text{v}_\text{M}=0.15\text{c}$ and $\alpha=0.7$ \citep[the $\text{v}_\text{M}$ and $\alpha$ values inferred for AT2017gfo,][]{waxman_constraints_2018}. Using the numerically calculated composition and the $\rho(\text{v})$ density profile, the semi-analytic approximations for $f_\gamma$ were obtained from eqs.~(\ref{eq:f_gamma_all_energies}), (\ref{eq:gamma_E_t_semi})-(\ref{eq:t_E_av}) and the analytic approximations were obtained from eqs.~\eqref{eq:f_geff_general}, \eqref{eq:t_geff_def}, \eqref{eq:k_geff_integral_def_2} and \eqref{eq:n_interpolation}. 
\footnotetext{This choice of $M$ ensures that the local $\rho t^3$ of most of the mass in the ejecta is similar to the $\rho t^3$ used to compute the composition.}

As in \S~\ref{sec:type_ia}, the MC calculations were carried out using URILIGHT (with the modifications of Appendix \ref{ap:URILIGHT}), and with $\Dot{Q}_\gamma(t)$, $\phi_\gamma(E,t)$ and $\gamma$-ray interaction opacities ($\kappa_{\textrm{PE}}$, $\kappa_{\textrm{C}}$ and $\kappa_{\textrm{PP}}$)\footnotemark calculated using the composition computed at the \textit{SkyNet} timestamp nearest to $t$.
\footnotetext{These (time dependent) opacities are well approximated (to a few percent accuracy near $t_\gamma$) by the opacities of a single element with atomic number $\approx Z_\text{eff}(t=t_\gamma)$ (see eq.~\refeq{eq:Z_eff}). This approximation gives $f_\gamma$ and $\Dot{Q}_{\text{dep}}$ with few percent and $\lesssim$1\% accuracy, respectively, at all relevant times.}

\begin{figure}
    \centering
    \includegraphics[width=\columnwidth]{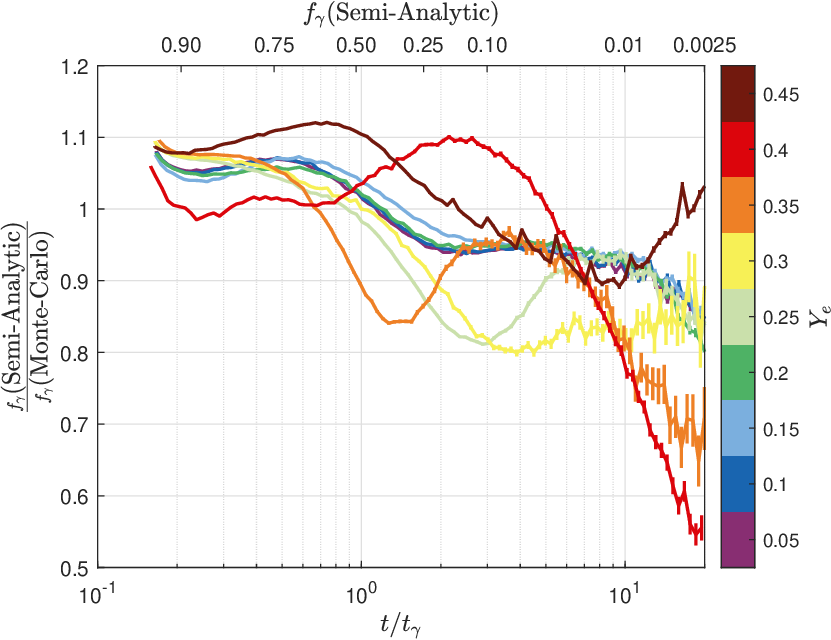}
    \\
    \includegraphics[width=\columnwidth]{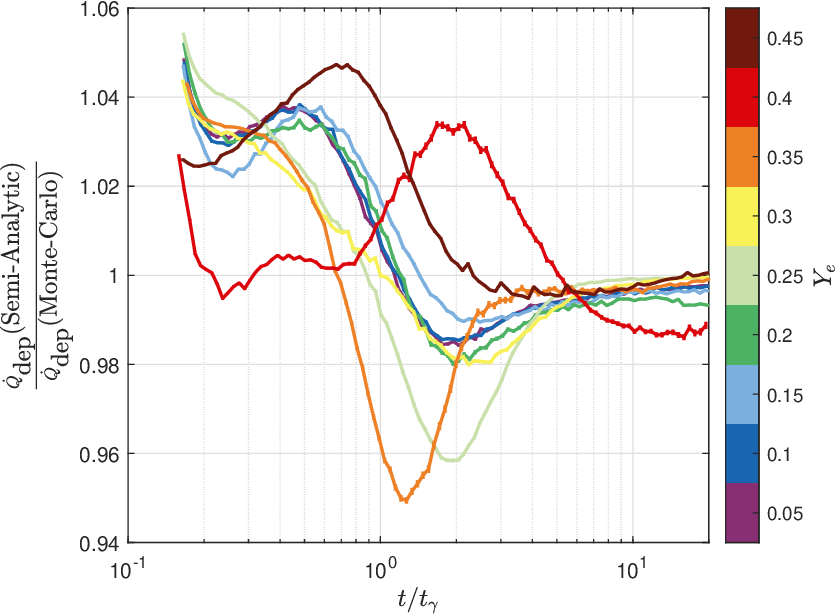}
    \caption{A comparison of the semi-analytic approximation, eqs.~(\ref{eq:f_gamma_all_energies}), (\ref{eq:gamma_E_t_semi})-(\ref{eq:t_E_av}), and the MC simulations results for $f_\gamma$ (\textit{Top}) and $\Dot{Q}_{\text{dep}}$ (\textit{Bottom}), for ejecta with $[\rho t^3]_{\text{KN}} = 1$ (eq. \ref{eq:rhoKN}), $s_0 = 20 \kb$ and various $Y_e$ values.
    The upper x-axis at the top panel shows the semi-analytic $f_\gamma$ values.
    The same interaction opacities are used in the URILIGHT MC calculations and in the semi-analytic approximation.
    The PE and PP opacities are adapted from XCOM data \citep{berger_xcom_2010}.
    The larger discrepancy in $f_\gamma$ at late times, when $f_\gamma<10^{-2}$, is mainly due to relativistic effects, which are not included in the semi-analytic and analytic approximations ("delayed" deposition of $\gamma$-rays accumulated in the ejecta from earlier times, that becomes significant at late times due to the decline in $\Dot{Q}_{\gamma}(t)$).} \label{fig:Semi_vs_MonteCarlo_KN}
\end{figure}
\begin{figure*}
     \includegraphics[width=\textwidth]{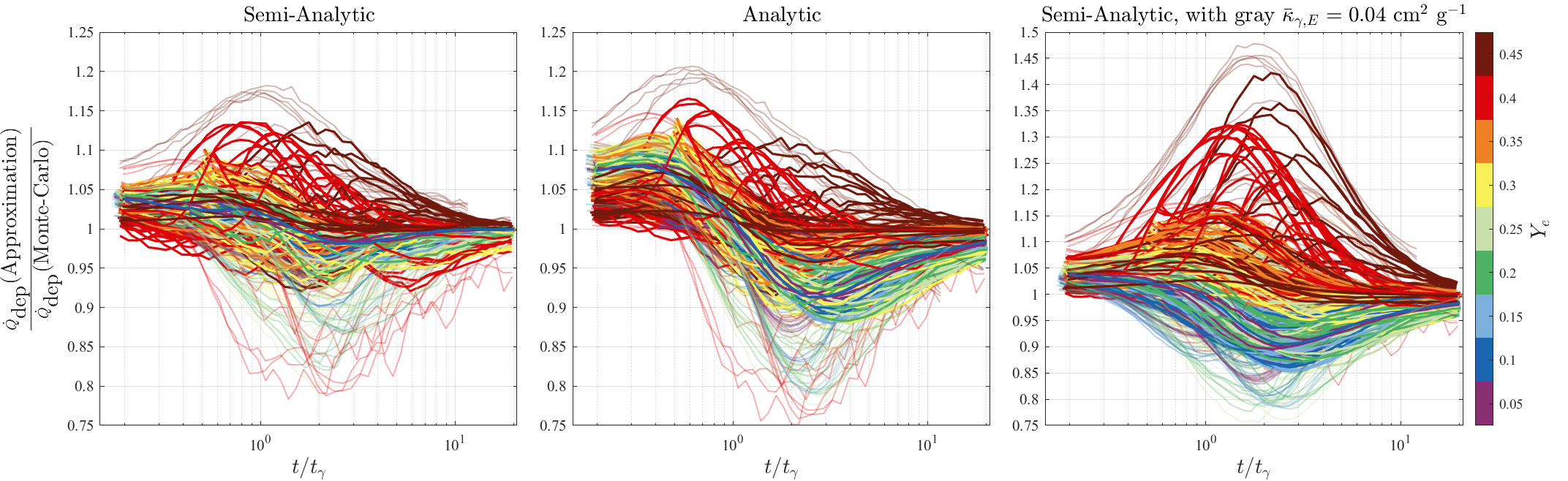}
    \caption{A comparison between $\Dot{Q}_{\rm dep}$ obtained from the MC simulations and those obtained by different approximations, for all ejecta parameter values explored in this work: \textit{Left} - the semi-analytic approximation, eqs.~(\ref{eq:f_gamma_all_energies}), (\ref{eq:gamma_E_t_semi})-(\ref{eq:t_E_av}); \textit{Middle} - the analytic approximation, eqs.~\eqref{eq:f_geff_general}, \eqref{eq:t_geff_def}, \eqref{eq:k_geff_integral_def_2} and \eqref{eq:n_interpolation}; \textit{Right} - the semi-analytic approximation, eqs.~(\ref{eq:f_gamma_all_energies}), (\ref{eq:gamma_E_t_semi})-(\ref{eq:t_E_av}), with fixed gray opacity $\Bar{\kappa}_{\gamma,E}=0.04$\cmgr.
     Semi-transparent thin lines, which show larger deviations of the approximations from MC results, indicate $[\rho t^3]_{\text{KN}}\geq10^2$ - $\rho t^3$ values much larger than those expected for NSM ejecta (see eq.~\refeq{eq:rhoKN}). For these cases, $t_\gamma\gtrsim10$~d and at such late times the decline in $\Dot{Q}_{\gamma}(t)$ results in larger discrepancies in $f_\gamma$ due to the relativistic effect of "delayed" deposition of $\gamma$-rays (see Fig. \ref{fig:Semi_vs_MonteCarlo_KN}).}
 \label{fig:dep_approx_compared_to_exact}
\end{figure*}
\begin{figure*}
 \includegraphics[width=\textwidth]{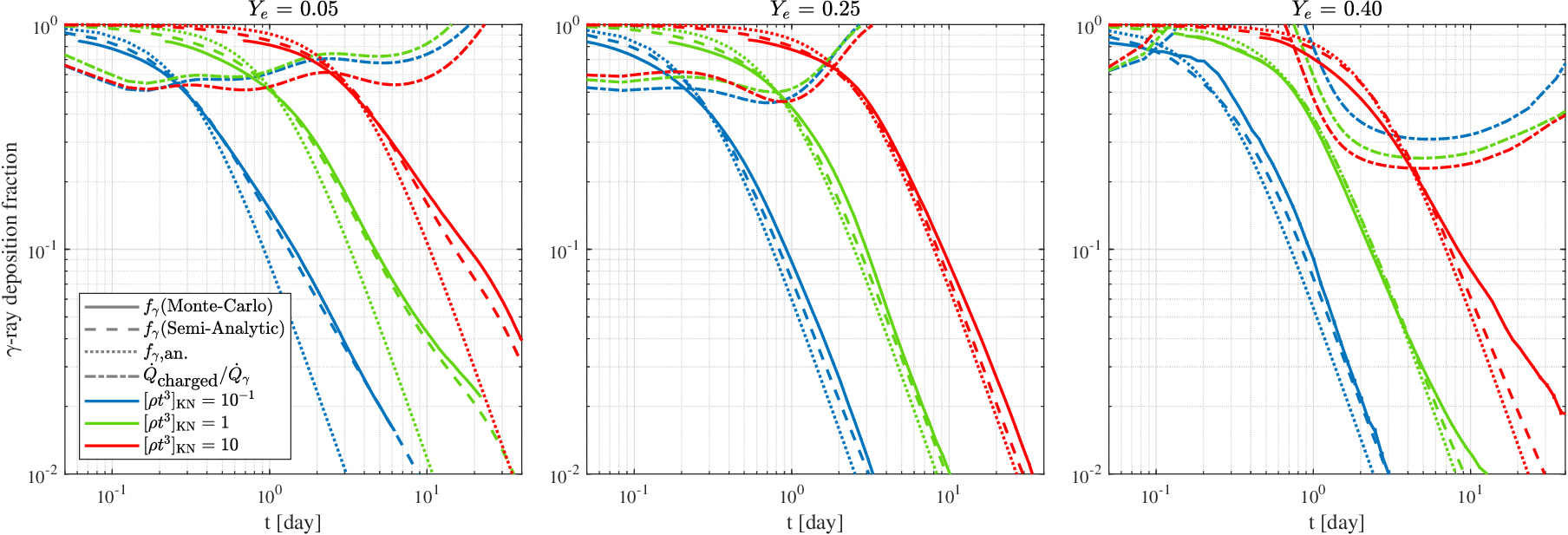}
     \caption{$f_\gamma(t)$ for KNe with $s_0 = 20 \kb$, and various $Y_e$, $[\rho t^3]_{\text{KN}}$ (eq. \ref{eq:rhoKN}) values. Solid, dashed, and dotted lines show the MC results, the semi-analytic approximation (eqs. \ref{eq:f_gamma_all_energies}, \ref{eq:gamma_E_t_semi}-\ref{eq:t_E_av}), and the analytic approximation (eq.~\ref{eq:f_geff_general} with $t_\gamma$ given by eqs. \ref{eq:t_geff_def} and~\ref{eq:k_geff_integral_def_2}, and $n=1.9$ as determined by eq. \ref{eq:n_interpolation}).
    Dash-dotted lines show $\Dot{Q}_{\text{charged}}/\Dot{Q}_{\gamma}$. As $f_\gamma$ drops below these lines, charged particles start to dominate the energy deposition in the ejecta, and inaccuracies in $f_\gamma$ do not affect significantly the inferred $\Dot{Q}_{\rm dep}$. The shallower than $t^{-2}$ decline of $f_\gamma$ at late time for low $Y_e$ is due to the extended deposition of energy by low-energy $\gamma$-rays in the high-$Z$ ejecta (the asymptotic $t^{-2}$ behavior will be reached at still later times).}
 \label{fig:kilonova_dep_frac}
\end{figure*}

Fig. \ref{fig:Semi_vs_MonteCarlo_KN} compares the semi-analytic and MC results for $f_\gamma$ (top) and $\Dot{Q}_{\text{dep}}$ (bottom), for ejecta with $[\rho t^3]_{\text{KN}}=1$ and $s_0 = 20~\kb$.

The left and middle panels of Fig. \ref{fig:dep_approx_compared_to_exact} compare $\Dot{Q}_{\text{dep}}$ obtained by the semi-analytic and analytic approximations, respectively, with that obtained by MC calculations, for all ejecta parameter values considered in this work. The right panel of Fig. \ref{fig:dep_approx_compared_to_exact} shows a similar comparison to the semi-analytic solution, eqs.~\eqref{eq:gamma_E_t_semi}-\eqref{eq:t_E_av}, obtained using fixed gray $\gamma$-ray opacity $\Bar{\kappa}_{\gamma,E}=0.04$\cmgr (motivated by $\kappa_\geff$ of eqs. \ref{eq:kilonova_kappa_approx}, \ref{eq:kilonova_Ye_th}).
Finally, Fig. \ref{fig:kilonova_dep_frac} compares $f_\gamma$ obtained from the MC simulations to our semi-analytic and analytic approximations, for various $Y_e$ values with $[\rho t^3]_{\text{KN}}=1$ and $s_0 = 20~\kb$.

Several conclusions may be drawn from the presented comparisons of MC and (semi)-analytic results.
\begin{enumerate}
    \item The semi-analytic expressions provide an excellent approximation to the MC results, with a 10--20\% accuracy in $f_\gamma$ and a few \% accuracy in $\Dot{Q}_{\rm dep}$\footnote{The slightly larger, 10\% accuracy for $\Dot{Q}_{\rm dep}$ at some $Y_e \geq 0.4$ cases is due to the combined effect of larger $\Dot{Q}_{\gamma}/\Dot{Q}_{\text{charged}}$ and 20\% deviation in $f_\gamma$.}. This demonstrates the validity of the approximations yielding the semi-analytic expressions (see \S~\ref{subsec:derivations}) - approximating the $\gamma$-ray transport by an appropriate absorption along the emission direction path, and neglecting relativistic effects (which become important for $f_\gamma$ only at late time, when $f_\gamma<10^{-2}$, and do not affect significantly $\Dot{Q}_{\rm dep}$, see Fig.~\ref{fig:Semi_vs_MonteCarlo_KN}; the same effect also leads to the worse accuracy, $\sim$20\% for dense ejecta with $[\rho t^3]_{\text{KN}}\geq10^2$ - values much larger than those expected for NSM, see Fig.~\ref{fig:dep_approx_compared_to_exact}). Since the approximations do not involve any assumptions regarding the spatial density and composition structure of the ejecta, we expect them, and hence the semi-analytic expressions, to hold for ejecta of general structure.  
    \item The analytic expressions provide an excellent, 10\% accuracy, approximation to the MC results for $\Dot{Q}_{\rm dep}$ in spherically symmetric ejecta with a wide range of density profiles. 
    \item Significant deviations of the analytic approximation from the MC results for $f_\gamma$ are obtained at late times and low $Y_e$ due to the extended deposition of energy by low-energy $\gamma$-rays in the high-$Z$ ejecta, see Fig. \ref{fig:kilonova_dep_frac}. This does not, however, affect
    $\Dot{Q}_{\rm dep}$ significantly due to the dominance of energy deposition by charged particles at these late times.
    \item The simplified semi-analytic description, with components-averaged $\Bar{\kappa}_{\gamma,E}$ (eqs.~\ref{eq:gamma_E_t_semi}-\ref{eq:t_E_av}), provides a very good approximation (introducing an increase in the inaccuracy of $\lesssim$10\%) for NSM ejecta, even under wide spatial variations of $Y_e$, the atomic composition and $\kappa_{\gamma,E}$. Consequently, the inaccuracy of the analytic approximation (defined to agree with the simplified semi-analytic solution at $t\sim t_\gamma$) is increased by the same level for such ejecta structures. This is shown by the 10\%--20\% increase in inaccuracy introduced by using a fixed value, $\Bar{\kappa}_{\gamma,E}=0.04$\cmgr for all ejecta parameters, in the semi-analytic approximation (the larger increases occur for the very neutron-poor cases of $Y_e\geq0.4$ and low $s_0$), a result in accordance with the estimate in \S~\ref{subsec:derivations}. For ejecta with spatially varying opacity, where the opacity is larger than $\Bar{\kappa}_{\gamma,E}$ in some regions and smaller in others, we expect the inaccuracy to be smaller than obtained for the cases examined here numerically, where the opacity is everywhere larger (or smaller) than $\Bar{\kappa}_{\gamma,E}$.
\end{enumerate}

\subsection{Comparison to earlier works}
\label{sec:comp_others}

We compare here our results to those of earlier works.
We note that in general, our results are based on nucleosynthesis calculations carried over a very wide range of ejecta parameter values, while earlier works focused on a narrower parameter value range \citep[e.g.][who focused on examining larger variations in the nuclear mass models]{barnes_kilonovae_2021}, or used other methods to derive the ejecta composition \citep[e.g.][who used a "backpropagated" solar (or solar-like) abundance]{hotokezaka_radioactive_2016,hotokezaka_radioactive_2020}.

Let us consider first past analytic estimates. Multiple earlier works used $\kappa_\geff\approx0.025$\cmgr, consistent with our results \citep[e.g.][]{metzger_electromagnetic_2010,hotokezaka_radioactive_2016,kasen_radioactive_2019}. These values were obtained qualitatively by correctly noting that most of the energy is carried by $\sim$1 MeV $\gamma$-rays and so $\kappa_\geff$ is dominated by Compton scattering, or by adopting the known results for Ia SNe. By using a quantitative method to derive $\kappa_\geff$ (for general $\gamma$-ray source spectra and plasma opacities), our work gives a firm basis for the former assumption ($\sim$1~MeV dominance), and allows examining the ejecta properties dependence of $\kappa_\geff$.

Other studies used the energy-weighted average opacity $\langle\kappa_{\gamma,E}\rangle$ (eq. \ref{eq:f_gamma_late}) as an estimate for the effective $\gamma$-ray opacity, leading to strong dependence of $\kappa_\geff$ on $Y_e$, with values larger by 1-2 orders of magnitude than the effective Compton scattering opacity at low $Y_e$. For example, \citet{barnes_radioactivity_2016} found $\langle\kappa_{\gamma,E}\rangle \approx 0.1$\cmgr for their low-$Y_e$ ejecta model \citep[a value later used for analytic or gray transfer calculations, e.g.][]{rosswog_detectability_2017, wollaeger_impact_2018,bulla_critical_2023};
\citet{hotokezaka_radioactive_2020} obtained $\langle\kappa_{\gamma,E}\rangle \approx 0.4(0.07)$\cmgr for an ejecta with heavy (light) r-process elements;
\citet{barnes_kilonovae_2021} obtained $\langle\kappa_{\gamma,E}\rangle$$\sim$1-3\cmgr for several nuclear mass model and low-$Y_e$ ejecta.
As discussed in \S~\ref{sec:intro}, and demonstrated explicitly for Ia SNe and KNe (in \S~\ref{sec:type_ia} and \S~\ref{subsec:KNe_kappa_geff}), $\langle\kappa_{\gamma,E}\rangle$ overestimates the contribution to the deposition of low-energy (high-opacity) $\gamma$-rays, and may therefore lead to a large overestimate of $f_\gamma$, $\kappa_\geff$ and $t_\gamma$, especially for high-$Z$ (low-$Y_e$) ejecta, where large PE absorption leads to higher opacities at low $\gamma$-ray energies. 

Let us discuss next earlier detailed numeric calculations. \citet{wu_fingerprints_2019} provided fits of $f_\gamma$ obtained by \citet{barnes_radioactivity_2016} (using MC calculations), $f_\gamma(t) \approx 1-e^{-t_\gamma^{1.28}/t^{1.28}}$, with  a list of mass-dependent $t_\gamma$, $t_\gamma(M)$.
Their values of $t_\gamma(M)$ correspond to $\kappa_\geff \approx 0.042$\cmgr (the column density of their model is $\langle\Sigma\rangle t^2\approx 10^{12} {\text{g}}\ {\text{cm}}^{-2}\ \text{s}^{2}$). Considering the ejecta of \citet{barnes_radioactivity_2016} is of low $Y_e$, this is consistent with our $\kappa_\geff$ at $Y_e\lesssim0.25$.
Moreover, their shallow late power law ($t^{-1.28}$ instead of $t^{-2}$) is consistent with the extended deposition of low-energy $\gamma$-rays in high-$Z$ ejecta (see Fig. \ref{fig:kilonova_dep_frac}), which is captured by the semi-analytic approximation
(see discussion in \S~\ref{subsec:comp_numerical}).

\citet{barnes_kilonovae_2021} conducted MC simulations to calculate $f_\gamma$ of KNe over the parameter range $0.16\leq Y_e\leq0.28$, $s_0 = 40 \kb$ and a  variety of nuclear mass models (using the density profile of \citet{barnes_effect_2013}, with $M=0.04M_\odot$, $\text{v}_\text{char}=0.1\text{c}$).
Their figure 4 shows the various $f_\gamma(t)$ obtained. While the late time behaviour of $f_\gamma$ differs considerably between different calculations (as expected, due to different low-energy $\gamma$-ray spectra, for which $\kappa_{\gamma,E}$ is affected by the different atomic compositions), $f_\gamma$ of all calculations are similar at $f_\gamma \gtrsim 10^{-0.5}$. This implies that $\kappa_\geff$ and $t_\gamma$ are similar in all calculations, despite the large differences in the ejecta parameters and nuclear mass models, a result consistent with our conclusions in \S~\ref{subsec:KNe_kappa_geff} and \S~\ref{subsec:nuclear_sensitivity}.

Finally, an examination of the $\gamma$-ray spectra reported by \citet{barnes_kilonovae_2021} is given in Appendix \ref{ap:Barnes}, showing that: (i) All their simulated $\gamma$-ray spectra, computed with a variety of nuclear mass models, show that most of the $\gamma$-ray energy is carried by $\sim$1~MeV photons, in agreement with our results;
(ii) $\kappa_\geff$ values much higher than we obtain, $\kappa_\geff \gg 0.03-0.05$\cmgr (e.g. $\kappa_\geff\sim1$\cmgr), require $\gamma$-ray spectra that are much softer than obtained in numeric simulations (as a large fraction of energy is required to be absorbed with a large opacity);
(iii) For the same nuclear mass model, there is a good agreement between our simulated spectra and those of \citet{barnes_kilonovae_2021}, despite the fact that they were computed using different nuclear reaction network codes.

\section{Summary}
\label{sec:summary}

We have derived a semi-analytic description, eqs. \eqref{eq:f_gamma_all_energies}, \eqref{eq:avg_f_gamma_angles}-\eqref{eq:multi-Sigma}, for $f_\gamma(t)$, the fraction of energy deposited by radioactive decay $\gamma$-rays, in a homologously expanding ejecta of general density distribution and atomic composition structure (spatial non-uniform $\gamma$-ray opacities $\kappa_{\gamma,E}(t)$).
Simplified approximate (exact for uniform $\kappa_{\gamma,E}$) expressions for $f_\gamma$, which use spatially averaged opacities $\Bar{\kappa}_{\gamma,E}$, are given by \eqref{eq:f_gamma_all_energies}, \eqref{eq:gamma_E_t_semi}-\eqref{eq:t_E_av}.
For spherically symmetric density distributions, an analytic approximation was derived, eqs.~\eqref{eq:f_geff_general}, \eqref{eq:t_geff_def}, \eqref{eq:k_geff_integral_def_2} and \eqref{eq:n_interpolation}.
For ejecta with spatial variations of the isotopic composition of the radioactive material (i.e. spatially non-uniform $\gamma$-ray spectrum $\phi_\gamma(E,t)$), $f_\gamma$ should be calculated separately for each radioactive component and the energy deposition is obtained by a sum over all components, $\sum_i f_{i\gamma}\Dot{Q}_{i\gamma}$.

The semi-analytic and analytic descriptions provide an excellent approximation for $f_\gamma$, reproducing with a few percent and 10\% accuracy, respectively, the energy deposition rate $\Dot{Q}_{\rm dep}$ obtained from MC calculations of $\gamma$-ray deposition. This was shown for well studied Ia SNe ejecta models (Figs. \ref{fig:typeIa_dep}, \ref{fig:Semi_vs_MonteCarlo_IA}) and for NSM ejecta models of various spherically symmetric density profiles and with a wide range of ejecta parameters $\{Y_e,s_0,\rho t^3\}$ and column densities (Figs. \ref{fig:Semi_vs_MonteCarlo_KN}, \ref{fig:dep_approx_compared_to_exact}, \ref{fig:kilonova_dep_frac}). While we examined here spherical ejecta of uniform atomic composition, the results demonstrate the applicability of the semi-analytic description for ejecta of general structure and of the analytic approximation for spherical ejecta of general structure (see discussion in \S~\ref{subsec:comp_numerical}):
\begin{itemize}
    \item The excellent agreement of the semi-analytic results with the results of MC calculations demonstrates the validity of the approximations yielding the semi-analytic expressions (see \S~\ref{subsec:derivations}) - approximating the $\gamma$-ray transport by an appropriate absorption along the emission direction path, and neglecting relativistic effects (which are negligible for Ia SNe and become important for KNe only at late time, when $f_\gamma<10^{-2}$, see Fig.~\ref{fig:Semi_vs_MonteCarlo_KN}). Since the approximations do not involve any assumptions regarding the spatial density and composition structure of the ejecta, we expect them, and hence the semi-analytic expressions, to hold for ejecta of general structure.
    \item For multi-component NSM ejecta, the use of spatially averaged opacities $\Bar{\kappa}_{\gamma,E}$ in the simplified semi-analytic solution (eqs.~\ref{eq:gamma_E_t_semi}-\ref{eq:t_E_av}) and the analytic approximation provides a very good approximation, introducing $\lesssim$10\% increase in inaccuracy, even for large spatial variations of $Y_e$ and atomic composition (Fig. \ref{fig:dep_approx_compared_to_exact}). This is well within the uncertainties in the radioactive energy production rate due to nuclear physics uncertainties.
\end{itemize}

We have shown that the effective $\gamma$-ray opacity, $\kappa_\geff$, that determines $t_\gamma$ through $t_\gamma=\sqrt{\kappa_\geff\langle\Sigma\rangle t^2}$, is given by eq.~\eqref{eq:k_geff_integral_def_2}. This definition yields the correct result, properly accounting for the contributions of both X-rays and $\gamma$-rays without an arbitrary omission of low-energy contributions (as commonly done in Ia SNe analyses, see e.g. Fig.~\ref{fig:typeIa_opacity}). For $\beta$-decay dominated energy release, $\kappa_\geff$ is typically close to the effective Compton scattering opacity, $\kappa_\geff \approx 0.025$\cmgr with a weak dependence on composition. For KNe, $\kappa_\geff$ depends mainly on the initial electron fraction $Y_e$,
$\kappa_\geff \approx 0.03(0.05)$\cmgr for $Y_e \gtrsim (\lesssim) 0.25$ (see eqs. \ref{eq:kilonova_kappa_approx}, \ref{eq:kilonova_Ye_th}) and Figs. \ref{fig:kilonova_kappa_eff_rhot3=1}, \ref{fig:kilonova_kappa_all_dependence}, \ref{fig:kilonova_kappa_eff_examination}), and is insensitive to the (large) nuclear physics uncertainties (see Fig.~\ref{fig:kilonova_random_rates_and_mass_model}).
This result, which is in contrast with some earlier work using $\kappa_\geff=\langle \kappa_{\gamma,E}\rangle$ yielding $\kappa_\geff$ larger than 0.025\cmgr by 1--2 orders of magnitude for low $Y_e$, is due to the fact that most of the $\gamma$-ray energy is carried by $\sim$1~MeV photons for which the opacity is insensitive to the exact atomic composition and $\gamma$-ray spectral shape. The reason for the large overestimate of the opacity obtained by using $\kappa_\geff=\langle \kappa_{\gamma,E}\rangle$ was explained in \S~\ref{sec:intro} (and demonstrated explicitly for Ia SNe and KNe in Figs.~\ref{fig:typeIa_opacity} and \ref{fig:kilonova_kappa_gamma_mean}, respectively).

The weak dependence of $\kappa_\geff$ on parameters implies that determining $t_\gamma$ from observations will measure the ejecta $\langle\Sigma\rangle t^2$, providing a stringent test of KN models. For $\langle\Sigma\rangle t^2=2\times10^{11}~{\rm g\,{cm}^{-2}\,s^2}$, a typical value expected for KNe, $t_\gamma\approx1$~d  (eqs. \ref{eq:kilonova_t_geff}, \ref{eq:SigKN}; Fig. \ref{fig:kilonova_t_gamma_groups}).

The calculation of the semi-analytic and analytic approximations requires the time-dependent $\gamma$-ray opacities and spectra, which are determined by the atomic and isotopic compositions, as well as ejecta structure parameters ($\langle\Sigma\rangle$, $f_\xi$, $n$; eqs. \ref{eq:asig_def}, \ref{eq:gamma_E_t_semi}, \ref{eq:n_interpolation}) which are determined by the density distribution. For KNe, accurate simplified approximations may be obtained using $\kappa_\geff(Y_e, s_0)$ from eqs. (\ref{eq:kilonova_kappa_approx}, \ref{eq:kilonova_Ye_th}) for uniform ejecta, or by using $\kappa_\geff = 0.04$\cmgr for ejecta of mixed composition. The latter increases the inaccuracy in predicting $\Dot{Q}_{\rm dep}$ by $\lesssim10\%$ (Fig. \ref{fig:dep_approx_compared_to_exact}). For NSM ejecta models we found $n\approx2$ (used in the analytic description, see eq. \ref{eq:f_geff_general} and Table \ref{table:interpolation}). Note that for low $Y_e$, our analytic approximation for $f_\gamma(t)$ is inaccurate at late times $t\gg t_\gamma$ (Fig. \ref{fig:kilonova_dep_frac}), but this does not affect the accuracy of determining $\Dot{Q}_{\rm dep}$ since charged particles dominate the energy deposition at $t>t_\gamma$.

Our model for $\gamma$-ray deposition can be applied to any radioactive expanding ejecta, including, e.g., core-collapse supernovae and ejecta produced by neutron star - black hole mergers. As an example, some Fast Blue Optical Transients (FBOTs) may be powered by the radioactive decay of ${}^{56}$Ni, with a fast declining luminosity that suggests $t_\gamma\sim1$ d \citep{ofek_at_2021}. From Fig. \ref{fig:typeIa_opacity} we conclude that for such fast events, $\kappa_\geff = 0.033$\cmgr should be used, instead of $\kappa_\geff = 0.025$\cmgr used for the slower ($t_\gamma\sim40$ d) Ia SNe (this is due to $t_\gamma$ occurring during the decay of ${}^{56}$Ni, instead of ${}^{56}$Co).

\section*{Acknowledgements}

We thank Amir Sharon and Doron Kushnir for fruitful discussions and their assistance with the URILIGHT code.

\section*{Data Availability}
A code for the numerical calculation of the semi-analytic expressions, eqs. \eqref{eq:avg_f_gamma_angles}-\eqref{eq:t_E_av}, including $f_\gamma(E,t)$, $f_\xi$, 
and $\asig t_0^2$, as well as of $n$ used in the analytic approximation (eq. \ref{eq:n_interpolation}), for multi-component ejecta with general structures $\rho_0(\textbf{v})$ and $\rho_{0\rm rad}(\textbf{v})$ is available at \url{https://github.com/or-guttman/gamma-ray-deposition}.

Numerical data will be shared following a reasonable request to the corresponding author.



\bibliographystyle{mnras}
\bibliography{refs}

\begin{thebibliography}{}
\makeatletter
\relax
\def\mn@urlcharsother{\let\do\@makeother \do\$\do\&\do\#\do\^\do\_\do\%\do\~}
\def\mn@doi{\begingroup\mn@urlcharsother \@ifnextchar [ {\mn@doi@}
  {\mn@doi@[]}}
\def\mn@doi@[#1]#2{\def\@tempa{#1}\ifx\@tempa\@empty \href
  {http://dx.doi.org/#2} {doi:#2}\else \href {http://dx.doi.org/#2} {#1}\fi
  \endgroup}
\def\mn@eprint#1#2{\mn@eprint@#1:#2::\@nil}
\def\mn@eprint@arXiv#1{\href {http://arxiv.org/abs/#1} {{\tt arXiv:#1}}}
\def\mn@eprint@dblp#1{\href {http://dblp.uni-trier.de/rec/bibtex/#1.xml}
  {dblp:#1}}
\def\mn@eprint@#1:#2:#3:#4\@nil{\def\@tempa {#1}\def\@tempb {#2}\def\@tempc
  {#3}\ifx \@tempc \@empty \let \@tempc \@tempb \let \@tempb \@tempa \fi \ifx
  \@tempb \@empty \def\@tempb {arXiv}\fi \@ifundefined
  {mn@eprint@\@tempb}{\@tempb:\@tempc}{\expandafter \expandafter \csname
  mn@eprint@\@tempb\endcsname \expandafter{\@tempc}}}

\bibitem[\protect\citeauthoryear{Ambwani \& Sutherland}{Ambwani \&
  Sutherland}{1988}]{ambwani_gamma-ray_1988}
Ambwani K.,  Sutherland P.,  1988, \mn@doi [The Astrophysical Journal]
  {10.1086/166052}, 325, 820

\bibitem[\protect\citeauthoryear{Axelrod}{Axelrod}{1980}]{axelrod_late_1980}
Axelrod T.~S.,  1980, Technical Report UCRL-52994, Late time optical spectra
  from the /sup 56/{Ni} model for {Type} {I} supernovae, \url
  {https://www.osti.gov/biblio/7096402}.
Lawrence Livermore National Lab. (LLNL), Livermore, CA (United States),
  \mn@doi{10.2172/7096402}, \url {https://www.osti.gov/biblio/7096402}

\bibitem[\protect\citeauthoryear{Barnes \& Kasen}{Barnes \&
  Kasen}{2013}]{barnes_effect_2013}
Barnes J.,  Kasen D.,  2013, \mn@doi [The Astrophysical Journal]
  {10.1088/0004-637X/775/1/18}, 775, 18

\bibitem[\protect\citeauthoryear{Barnes, Kasen, Wu  \& Martínez-Pinedo}{Barnes
  et~al.}{2016}]{barnes_radioactivity_2016}
Barnes J.,  Kasen D.,  Wu M.-R.,   Martínez-Pinedo G.,  2016, \mn@doi [The
  Astrophysical Journal] {10.3847/0004-637X/829/2/110}, 829, 110

\bibitem[\protect\citeauthoryear{Barnes, Zhu, Lund, Sprouse, Vassh, McLaughlin,
  Mumpower  \& Surman}{Barnes et~al.}{2021}]{barnes_kilonovae_2021}
Barnes J.,  Zhu Y.~L.,  Lund K.~A.,  Sprouse T.~M.,  Vassh N.,  McLaughlin
  G.~C.,  Mumpower M.~R.,   Surman R.,  2021, \mn@doi [The Astrophysical
  Journal] {10.3847/1538-4357/ac0aec}, 918, 44

\bibitem[\protect\citeauthoryear{Berger, Hubbell, Seltzer, Chang, Coursey,
  Sukumar, Zucker  \& Olsen}{Berger et~al.}{2010}]{berger_xcom_2010}
Berger M.,  Hubbell J.,  Seltzer S.,  Chang J.,  Coursey J.,  Sukumar R.,
  Zucker D.,   Olsen K.,  2010, {XCOM}: {Photon} {Cross} {Sections} {Database},
  \url {https://www.nist.gov/pml/xcom-photon-cross-sections-database}

\bibitem[\protect\citeauthoryear{Blondin et~al.,}{Blondin
  et~al.}{2022}]{blondin_standart_2022}
Blondin S.,  et~al., 2022, \mn@doi [Astronomy and Astrophysics]
  {10.1051/0004-6361/202244134}, 668, A163

\bibitem[\protect\citeauthoryear{Brown et~al.,}{Brown
  et~al.}{2018}]{brown_endfb-viii0_2018}
Brown D.~A.,  et~al., 2018, \mn@doi [Nuclear Data Sheets]
  {10.1016/j.nds.2018.02.001}, 148, 1

\bibitem[\protect\citeauthoryear{Bulla}{Bulla}{2023}]{bulla_critical_2023}
Bulla M.,  2023, \mn@doi [Monthly Notices of the Royal Astronomical Society]
  {10.1093/mnras/stad232}, 520, 2558

\bibitem[\protect\citeauthoryear{Colgate \& McKee}{Colgate \&
  McKee}{1969}]{colgate_early_1969}
Colgate S.~A.,  McKee C.,  1969, \mn@doi [The Astrophysical Journal]
  {10.1086/150102}, 157, 623

\bibitem[\protect\citeauthoryear{Colgate, Petschek  \& Kriese}{Colgate
  et~al.}{1980}]{colgate_luminosity_1980}
Colgate S.~A.,  Petschek A.~G.,   Kriese J.~T.,  1980, \mn@doi [The
  Astrophysical Journal] {10.1086/183239}, 237, L81

\bibitem[\protect\citeauthoryear{Cowan, Sneden, Lawler, Aprahamian, Wiescher,
  Langanke, Martínez-Pinedo  \& Thielemann}{Cowan
  et~al.}{2021}]{cowan_origin_2021}
Cowan J.~J.,  Sneden C.,  Lawler J.~E.,  Aprahamian A.,  Wiescher M.,  Langanke
  K.,  Martínez-Pinedo G.,   Thielemann F.-K.,  2021, \mn@doi [Reviews of
  Modern Physics] {10.1103/RevModPhys.93.015002}, 93, 015002

\bibitem[\protect\citeauthoryear{Cyburt et~al.,}{Cyburt
  et~al.}{2010}]{cyburt_jina_2010}
Cyburt R.~H.,  et~al., 2010, \mn@doi [The Astrophysical Journal Supplement
  Series] {10.1088/0067-0049/189/1/240}, 189, 240

\bibitem[\protect\citeauthoryear{Drout et~al.,}{Drout
  et~al.}{2017}]{drout_light_2017}
Drout M.~R.,  et~al., 2017, \mn@doi [Science] {10.1126/science.aaq0049}, 358,
  1570

\bibitem[\protect\citeauthoryear{Erler, Birge, Kortelainen, Nazarewicz, Olsen,
  Perhac  \& Stoitsov}{Erler et~al.}{2012}]{erler_limits_2012}
Erler J.,  Birge N.,  Kortelainen M.,  Nazarewicz W.,  Olsen E.,  Perhac A.~M.,
    Stoitsov M.,  2012, \mn@doi [Nature] {10.1038/nature11188}, 486, 509

\bibitem[\protect\citeauthoryear{Farouqi, Kratz, Pfeiffer, Rauscher, Thielemann
   \& Truran}{Farouqi et~al.}{2010}]{farouqi_charged-particle_2010}
Farouqi K.,  Kratz K.-L.,  Pfeiffer B.,  Rauscher T.,  Thielemann F.-K.,
  Truran J.~W.,  2010, \mn@doi [The Astrophysical Journal]
  {10.1088/0004-637X/712/2/1359}, 712, 1359

\bibitem[\protect\citeauthoryear{Fernández \& Metzger}{Fernández \&
  Metzger}{2016}]{fernandez_electromagnetic_2016}
Fernández R.,  Metzger B.~D.,  2016, \mn@doi [Annual Review of Nuclear and
  Particle Science] {10.1146/annurev-nucl-102115-044819}, 66, 23

\bibitem[\protect\citeauthoryear{Freiburghaus, Rosswog  \&
  Thielemann}{Freiburghaus et~al.}{1999}]{freiburghaus_r-process_1999}
Freiburghaus C.,  Rosswog S.,   Thielemann F.-K.,  1999, \mn@doi [The
  Astrophysical Journal] {10.1086/312343}, 525, L121

\bibitem[\protect\citeauthoryear{Hotokezaka \& Nakar}{Hotokezaka \&
  Nakar}{2020}]{hotokezaka_radioactive_2020}
Hotokezaka K.,  Nakar E.,  2020, \mn@doi [The Astrophysical Journal]
  {10.3847/1538-4357/ab6a98}, 891, 152

\bibitem[\protect\citeauthoryear{Hotokezaka, Wanajo, Tanaka, Bamba, Terada  \&
  Piran}{Hotokezaka et~al.}{2016}]{hotokezaka_radioactive_2016}
Hotokezaka K.,  Wanajo S.,  Tanaka M.,  Bamba A.,  Terada Y.,   Piran T.,
  2016, \mn@doi [Monthly Notices of the Royal Astronomical Society]
  {10.1093/mnras/stw404}, 459, 35

\bibitem[\protect\citeauthoryear{Hotokezaka, Sari  \& Piran}{Hotokezaka
  et~al.}{2017}]{hotokezaka_analytic_2017}
Hotokezaka K.,  Sari R.,   Piran T.,  2017, \mn@doi [Monthly Notices of the
  Royal Astronomical Society] {10.1093/mnras/stx411}, 468, 91

\bibitem[\protect\citeauthoryear{Hotokezaka, Beniamini  \& Piran}{Hotokezaka
  et~al.}{2018}]{hotokezaka_neutron_2018}
Hotokezaka K.,  Beniamini P.,   Piran T.,  2018, \mn@doi [International Journal
  of Modern Physics D] {10.1142/S0218271818420051}, 27, 1842005

\bibitem[\protect\citeauthoryear{Jeffery}{Jeffery}{1999}]{jeffery_radioactive_1999}
Jeffery D.~J.,  1999, \mn@doi [arXiv e-prints]
  {10.48550/arXiv.astro-ph/9907015}, pp astro--ph/9907015

\bibitem[\protect\citeauthoryear{Kasen \& Barnes}{Kasen \&
  Barnes}{2019}]{kasen_radioactive_2019}
Kasen D.,  Barnes J.,  2019, \mn@doi [The Astrophysical Journal]
  {10.3847/1538-4357/ab06c2}, 876, 128

\bibitem[\protect\citeauthoryear{Kasen, Metzger, Barnes, Quataert  \&
  Ramirez-Ruiz}{Kasen et~al.}{2017}]{kasen_origin_2017}
Kasen D.,  Metzger B.,  Barnes J.,  Quataert E.,   Ramirez-Ruiz E.,  2017,
  \mn@doi [Nature] {10.1038/nature24453}, 551, 80

\bibitem[\protect\citeauthoryear{Kasliwal et~al.,}{Kasliwal
  et~al.}{2017}]{kasliwal_illuminating_2017}
Kasliwal M.~M.,  et~al., 2017, \mn@doi [Science] {10.1126/science.aap9455},
  358, 1559

\bibitem[\protect\citeauthoryear{Kortelainen, McDonnell, Nazarewicz, Reinhard,
  Sarich, Schunck, Stoitsov  \& Wild}{Kortelainen
  et~al.}{2012}]{kortelainen_nuclear_2012}
Kortelainen M.,  McDonnell J.,  Nazarewicz W.,  Reinhard P.~G.,  Sarich J.,
  Schunck N.,  Stoitsov M.~V.,   Wild S.~M.,  2012, \mn@doi [Physical Review C]
  {10.1103/PhysRevC.85.024304}, 85, 024304

\bibitem[\protect\citeauthoryear{Kullmann, Goriely, Just, Bauswein  \&
  Janka}{Kullmann et~al.}{2023}]{kullmann_impact_2023}
Kullmann I.,  Goriely S.,  Just O.,  Bauswein A.,   Janka H.~T.,  2023, \mn@doi
  [Monthly Notices of the Royal Astronomical Society] {10.1093/mnras/stad1458},
  523, 2551

\bibitem[\protect\citeauthoryear{Lattimer \& Schramm}{Lattimer \&
  Schramm}{1974}]{lattimer_black-hole-neutron-star_1974}
Lattimer J.~M.,  Schramm D.~N.,  1974, \mn@doi [The Astrophysical Journal]
  {10.1086/181612}, 192, L145

\bibitem[\protect\citeauthoryear{Li \& Paczyński}{Li \&
  Paczyński}{1998}]{li_transient_1998}
Li L.-X.,  Paczyński B.,  1998, \mn@doi [The Astrophysical Journal]
  {10.1086/311680}, 507, L59

\bibitem[\protect\citeauthoryear{Lippuner \& Roberts}{Lippuner \&
  Roberts}{2015}]{lippuner_r-process_2015}
Lippuner J.,  Roberts L.~F.,  2015, \mn@doi [The Astrophysical Journal]
  {10.1088/0004-637X/815/2/82}, 815, 82

\bibitem[\protect\citeauthoryear{Lippuner \& Roberts}{Lippuner \&
  Roberts}{2017}]{lippuner_skynet_2017}
Lippuner J.,  Roberts L.~F.,  2017, \mn@doi [The Astrophysical Journal
  Supplement Series] {10.3847/1538-4365/aa94cb}, 233, 18

\bibitem[\protect\citeauthoryear{Longair}{Longair}{2011}]{longair_high_2011}
Longair M.~S.,  2011, High {Energy} {Astrophysics}.
\url {https://ui.adsabs.harvard.edu/abs/2011hea..book.....L}

\bibitem[\protect\citeauthoryear{Lund, Engel, McLaughlin, Mumpower, Ney  \&
  Surman}{Lund et~al.}{2023}]{lund_influence_2023}
Lund K.~A.,  Engel J.,  McLaughlin G.~C.,  Mumpower M.~R.,  Ney E.~M.,   Surman
  R.,  2023, \mn@doi [The Astrophysical Journal] {10.3847/1538-4357/acaf56},
  944, 144

\bibitem[\protect\citeauthoryear{Maoz, Mannucci  \& Nelemans}{Maoz
  et~al.}{2014}]{maoz_observational_2014}
Maoz D.,  Mannucci F.,   Nelemans G.,  2014, \mn@doi [Annual Review of
  Astronomy and Astrophysics] {10.1146/annurev-astro-082812-141031}, 52, 107

\bibitem[\protect\citeauthoryear{Metzger}{Metzger}{2019}]{metzger_kilonovae_2019}
Metzger B.~D.,  2019, \mn@doi [Living Reviews in Relativity]
  {10.1007/s41114-019-0024-0}, 23, 1

\bibitem[\protect\citeauthoryear{Metzger et~al.,}{Metzger
  et~al.}{2010}]{metzger_electromagnetic_2010}
Metzger B.~D.,  et~al., 2010, \mn@doi [Monthly Notices of the Royal
  Astronomical Society] {10.1111/j.1365-2966.2010.16864.x}, 406, 2650

\bibitem[\protect\citeauthoryear{Meyer}{Meyer}{1993}]{meyer_entropy_1993}
Meyer B.~S.,  1993, \mn@doi [Physics Reports] {10.1016/0370-1573(93)90071-K},
  227, 257

\bibitem[\protect\citeauthoryear{Mougeot}{Mougeot}{2017}]{mougeot_betashape_2017}
Mougeot X.,  2017, \mn@doi [EPJ Web of Conferences]
  {10.1051/epjconf/201714612015}, 146, 12015

\bibitem[\protect\citeauthoryear{Mumpower, Cass, Passucci, Surman  \&
  Aprahamian}{Mumpower et~al.}{2014}]{mumpower_sensitivity_2014}
Mumpower M.,  Cass J.,  Passucci G.,  Surman R.,   Aprahamian A.,  2014,
  \mn@doi [AIP Advances] {10.1063/1.4867192}, 4, 041009

\bibitem[\protect\citeauthoryear{Mumpower, Surman, Fang, Beard, Möller, Kawano
   \& Aprahamian}{Mumpower et~al.}{2015}]{mumpower_impact_2015}
Mumpower M.~R.,  Surman R.,  Fang D.-L.,  Beard M.,  Möller P.,  Kawano T.,
  Aprahamian A.,  2015, \mn@doi [Physical Review C]
  {10.1103/PhysRevC.92.035807}, 92, 035807

\bibitem[\protect\citeauthoryear{Mumpower, Surman, McLaughlin  \&
  Aprahamian}{Mumpower et~al.}{2016}]{mumpower_impact_2016}
Mumpower M.~R.,  Surman R.,  McLaughlin G.~C.,   Aprahamian A.,  2016, \mn@doi
  [Progress in Particle and Nuclear Physics] {10.1016/j.ppnp.2015.09.001}, 86,
  86

\bibitem[\protect\citeauthoryear{Möller, Sierk, Ichikawa  \& Sagawa}{Möller
  et~al.}{2016}]{moller_nuclear_2016}
Möller P.,  Sierk A.~J.,  Ichikawa T.,   Sagawa H.,  2016, \mn@doi [Atomic
  Data and Nuclear Data Tables] {10.1016/j.adt.2015.10.002}, 109-110, 1

\bibitem[\protect\citeauthoryear{Nakar}{Nakar}{2020}]{nakar_electromagnetic_2020}
Nakar E.,  2020, \mn@doi [Physics Reports] {10.1016/j.physrep.2020.08.008},
  886, 1

\bibitem[\protect\citeauthoryear{Nedora et~al.,}{Nedora
  et~al.}{2021}]{nedora_numerical_2021}
Nedora V.,  et~al., 2021, \mn@doi [The Astrophysical Journal]
  {10.3847/1538-4357/abc9be}, 906, 98

\bibitem[\protect\citeauthoryear{Ofek et~al.,}{Ofek
  et~al.}{2021}]{ofek_at_2021}
Ofek E.~O.,  et~al., 2021, \mn@doi [The Astrophysical Journal]
  {10.3847/1538-4357/ac24fc}, 922, 247

\bibitem[\protect\citeauthoryear{Pankey}{Pankey}{1962}]{pankey_possible_1962}
Pankey Jr. T.,  1962, PhD thesis, \url
  {https://ui.adsabs.harvard.edu/abs/1962PhDT........25P}

\bibitem[\protect\citeauthoryear{Perego, Thielemann  \& Cescutti}{Perego
  et~al.}{2021}]{perego_r-process_2021}
Perego A.,  Thielemann F.~K.,   Cescutti G.,  2021, r-{Process}
  {Nucleosynthesis} from {Compact} {Binary} {Mergers},
  \mn@doi{10.1007/978-981-15-4702-7_13-1.
}, \url {https://ui.adsabs.harvard.edu/abs/2021hgwa.bookE..13P}

\bibitem[\protect\citeauthoryear{Perego et~al.,}{Perego
  et~al.}{2022}]{perego_production_2022}
Perego A.,  et~al., 2022, \mn@doi [The Astrophysical Journal]
  {10.3847/1538-4357/ac3751}, 925, 22

\bibitem[\protect\citeauthoryear{Radice, Galeazzi, Lippuner, Roberts, Ott  \&
  Rezzolla}{Radice et~al.}{2016}]{radice_dynamical_2016}
Radice D.,  Galeazzi F.,  Lippuner J.,  Roberts L.~F.,  Ott C.~D.,   Rezzolla
  L.,  2016, \mn@doi [Monthly Notices of the Royal Astronomical Society]
  {10.1093/mnras/stw1227}, 460, 3255

\bibitem[\protect\citeauthoryear{Radice, Perego, Hotokezaka, Fromm, Bernuzzi
  \& Roberts}{Radice et~al.}{2018}]{radice_binary_2018}
Radice D.,  Perego A.,  Hotokezaka K.,  Fromm S.~A.,  Bernuzzi S.,   Roberts
  L.~F.,  2018, \mn@doi [The Astrophysical Journal] {10.3847/1538-4357/aaf054},
  869, 130

\bibitem[\protect\citeauthoryear{Radice, Bernuzzi  \& Perego}{Radice
  et~al.}{2020}]{radice_dynamics_2020}
Radice D.,  Bernuzzi S.,   Perego A.,  2020, \mn@doi [Annual Review of Nuclear
  and Particle Science] {10.1146/annurev-nucl-013120-114541}, 70, 95

\bibitem[\protect\citeauthoryear{Roberts, Kasen, Lee  \& Ramirez-Ruiz}{Roberts
  et~al.}{2011}]{roberts_electromagnetic_2011}
Roberts L.~F.,  Kasen D.,  Lee W.~H.,   Ramirez-Ruiz E.,  2011, \mn@doi [The
  Astrophysical Journal Letters] {10.1088/2041-8205/736/1/L21}, 736, L21

\bibitem[\protect\citeauthoryear{Rosswog \& Korobkin}{Rosswog \&
  Korobkin}{2024}]{rosswog_heavy_2024}
Rosswog S.,  Korobkin O.,  2024, \mn@doi [Annalen der Physik]
  {10.1002/andp.202200306}, 536, 2200306

\bibitem[\protect\citeauthoryear{Rosswog, Feindt, Korobkin, Wu, Sollerman,
  Goobar  \& Martinez-Pinedo}{Rosswog
  et~al.}{2017}]{rosswog_detectability_2017}
Rosswog S.,  Feindt U.,  Korobkin O.,  Wu M.~R.,  Sollerman J.,  Goobar A.,
  Martinez-Pinedo G.,  2017, \mn@doi [Classical and Quantum Gravity]
  {10.1088/1361-6382/aa68a9}, 34, 104001

\bibitem[\protect\citeauthoryear{Scalzo et~al.,}{Scalzo
  et~al.}{2014}]{scalzo_type_2014}
Scalzo R.,  et~al., 2014, \mn@doi [Monthly Notices of the Royal Astronomical
  Society] {10.1093/mnras/stu350}, 440, 1498

\bibitem[\protect\citeauthoryear{Sharon \& Kushnir}{Sharon \&
  Kushnir}{2020}]{sharon_-ray_2020}
Sharon A.,  Kushnir D.,  2020, \mn@doi [Monthly Notices of the Royal
  Astronomical Society] {10.1093/mnras/staa1745}, 496, 4517

\bibitem[\protect\citeauthoryear{Shenhar, Guttman  \& Waxman}{Shenhar
  et~al.}{2024}]{shenhar_analytic_2024}
Shenhar B.,  Guttman O.,   Waxman E.,  2024, An {Analytic} {Description} of
  {Electron} {Thermalization} in {Kilonovae} {Ejecta}

\bibitem[\protect\citeauthoryear{Shibata \& Hotokezaka}{Shibata \&
  Hotokezaka}{2019}]{shibata_merger_2019}
Shibata M.,  Hotokezaka K.,  2019, \mn@doi [Annual Review of Nuclear and
  Particle Science] {10.1146/annurev-nucl-101918-023625}, 69, 41

\bibitem[\protect\citeauthoryear{Shultis \& Faw}{Shultis \&
  Faw}{2000}]{shultis_radiation_2000}
Shultis J.,  Faw R.,  2000, Radiation {Shielding}.
American Nuclear Society, \url
  {https://books.google.co.il/books?id=HiN3QgAACAAJ}

\bibitem[\protect\citeauthoryear{Surman, Beun, McLaughlin  \& Hix}{Surman
  et~al.}{2009}]{surman_neutron_2009}
Surman R.,  Beun J.,  McLaughlin G.~C.,   Hix W.~R.,  2009, \mn@doi [Physical
  Review C] {10.1103/PhysRevC.79.045809}, 79, 045809

\bibitem[\protect\citeauthoryear{Surman, Mumpower, Cass, Bentley, Aprahamian
  \& McLaughlin}{Surman et~al.}{2014}]{surman_sensitivity_2014}
Surman R.,  Mumpower M.,  Cass J.,  Bentley I.,  Aprahamian A.,   McLaughlin
  G.~C.,  2014, \mn@doi [EPJ Web of Conferences] {10.1051/epjconf/20146607024},
  66, 07024

\bibitem[\protect\citeauthoryear{Sutherland \& Wheeler}{Sutherland \&
  Wheeler}{1984}]{sutherland_models_1984}
Sutherland P.~G.,  Wheeler J.~C.,  1984, \mn@doi [The Astrophysical Journal]
  {10.1086/161995}, 280, 282

\bibitem[\protect\citeauthoryear{Swartz, Sutherland  \& Harkness}{Swartz
  et~al.}{1995}]{swartz_gamma-ray_1995}
Swartz D.~A.,  Sutherland P.~G.,   Harkness R.~P.,  1995, \mn@doi [The
  Astrophysical Journal] {10.1086/175834}, 446, 766

\bibitem[\protect\citeauthoryear{Symbalisty \& Schramm}{Symbalisty \&
  Schramm}{1982}]{symbalisty_neutron_1982}
Symbalisty E.,  Schramm D.~N.,  1982, Astrophysical Letters, 22, 143

\bibitem[\protect\citeauthoryear{Waxman, Ofek, Kushnir  \& Gal-Yam}{Waxman
  et~al.}{2018}]{waxman_constraints_2018}
Waxman E.,  Ofek E.,  Kushnir D.,   Gal-Yam A.,  2018, \mn@doi [Monthly Notices
  of the Royal Astronomical Society] {10.1093/mnras/sty2441}, 481, 3423

\bibitem[\protect\citeauthoryear{Waxman, Ofek  \& Kushnir}{Waxman
  et~al.}{2019}]{waxman_late-time_2019}
Waxman E.,  Ofek E.~O.,   Kushnir D.,  2019, \mn@doi [The Astrophysical
  Journal] {10.3847/1538-4357/ab1f71}, 878, 93

\bibitem[\protect\citeauthoryear{Way \& Wigner}{Way \&
  Wigner}{1948}]{way_rate_1948}
Way K.,  Wigner E.~P.,  1948, \mn@doi [Physical Review]
  {10.1103/PhysRev.73.1318}, 73, 1318

\bibitem[\protect\citeauthoryear{Weaver, Axelrod  \& Woosley}{Weaver
  et~al.}{1980}]{weaver_type_1980}
Weaver T.~A.,  Axelrod T.~S.,   Woosley S.~E.,  1980, Technical Report
  UCRL-85121; CONF-800370-3, Type {I} supernova models vs observations, \url
  {https://www.osti.gov/biblio/6756382}.
California Univ., Livermore (USA). Lawrence Livermore National Lab.; California
  Univ., Santa Cruz (USA). Lick Observatory, \url
  {https://www.osti.gov/biblio/6756382}

\bibitem[\protect\citeauthoryear{Wilk, Hillier  \& Dessart}{Wilk
  et~al.}{2019}]{wilk_solving_2019}
Wilk K.~D.,  Hillier D.~J.,   Dessart L.,  2019, \mn@doi [Monthly Notices of
  the Royal Astronomical Society] {10.1093/mnras/stz1367}, 487, 1218

\bibitem[\protect\citeauthoryear{Wollaeger et~al.,}{Wollaeger
  et~al.}{2018}]{wollaeger_impact_2018}
Wollaeger R.~T.,  et~al., 2018, \mn@doi [Monthly Notices of the Royal
  Astronomical Society] {10.1093/mnras/sty1018}, 478, 3298

\bibitem[\protect\citeauthoryear{Wu, Barnes, Martínez-Pinedo  \& Metzger}{Wu
  et~al.}{2019}]{wu_fingerprints_2019}
Wu M.-R.,  Barnes J.,  Martínez-Pinedo G.,   Metzger B.~D.,  2019, \mn@doi
  [Physical Review Letters] {10.1103/PhysRevLett.122.062701}, 122, 062701

\bibitem[\protect\citeauthoryear{Wygoda, Elbaz  \& Katz}{Wygoda
  et~al.}{2019a}]{wygoda_type_2019-1}
Wygoda N.,  Elbaz Y.,   Katz B.,  2019a, \mn@doi [Monthly Notices of the Royal
  Astronomical Society] {10.1093/mnras/stz145}, 484, 3941

\bibitem[\protect\citeauthoryear{Wygoda, Elbaz  \& Katz}{Wygoda
  et~al.}{2019b}]{wygoda_type_2019}
Wygoda N.,  Elbaz Y.,   Katz B.,  2019b, \mn@doi [Monthly Notices of the Royal
  Astronomical Society] {10.1093/mnras/stz146}, 484, 3951

\bibitem[\protect\citeauthoryear{Zhu, Lund, Barnes, Sprouse, Vassh, McLaughlin,
  Mumpower  \& Surman}{Zhu et~al.}{2021}]{zhu_modeling_2021}
Zhu Y.~L.,  Lund K.~A.,  Barnes J.,  Sprouse T.~M.,  Vassh N.,  McLaughlin
  G.~C.,  Mumpower M.~R.,   Surman R.,  2021, \mn@doi [The Astrophysical
  Journal] {10.3847/1538-4357/abc69e}, 906, 94

\makeatother
\end{thebibliography}



\appendix

\section{The $\gamma$-ray spectrum for which $\kappa_\geff \gg 0.025$\cmgr~ may be obtained}
\label{ap:g-spec-large-kappa}
Here we derive stringent constraints on the $\gamma$-ray spectrum which may yield $\kappa_\geff \gg 0.025$\cmgr, and show that such values require $\gamma$-ray spectra much softer than expected in general for $\beta$-decay dominated energy release (typically dominated by $\sim$1 MeV photons).
We consider $\phi_\gamma$ as a probability distribution, and derive Chernoff bounds from eq. \eqref{eq:k_geff_integral_def_2}, for the fraction $x$ ($0\leq x\leq1$) of $\gamma$-ray energy carried by $\gamma$-rays which have $\kappa_{\gamma,E}$ greater than some value that depends on $\kappa_\geff$,
\begin{equation}
    P\left(\kappa_{\gamma,E} \geq [1+\ln(1-x)] \kappa_\geff\right) \geq x.
    \label{eq:chernoff_kappa}
\end{equation}
For example, substituting $x=0.5$ implies that at least half the $\gamma$-ray energy is emitted as $\gamma$-rays with $\kappa_{\gamma,E}\geq0.31\kappa_\geff$.
Using $\kappa_{\gamma,E}$ this inequality can be converted to a constraint on $\phi_\gamma$.
A crude analytic constraint on the (energy-weighted) spectrum is
$P\left(E_{\text{MeV}} \lesssim \frac{Z}{400}([1+\ln(1-x)]\kappa_\geff)^{-0.4}\right)\gtrsim x$, where $\kappa_\geff$ is in \cmgr and $E_{\text{MeV}}$ is in MeV, derived by taking PE dominated $\kappa_{\gamma,E}$, that is very crudely $\kappa_{\gamma,E} \sim (\frac{1}{400}\frac{Z}{E_{\text{MeV}}})^{2.5}$\cmgr.

Fig. \ref{fig:chernoff_E_med} shows upper bounds on the medians of the energy-weighted $\gamma$-ray spectra, as a function of $\kappa_\geff$ for different materials.
For high values of $\kappa_\geff$, the $\gamma$-ray spectrum is restricted to be considerably soft, e.g. $\kappa_\geff \sim 1$\cmgr requires that half of the energy in $\gamma$-rays is emitted in $<$300 keV photons even for high-$Z$ ejecta. Similar bounds can be calculated for any percentile, yielding a continuous and stringent constraint on the $\gamma$-ray spectrum (as shown in Appendix \ref{ap:Barnes}).

\begin{figure}
\centering
 \includegraphics[width=\columnwidth]{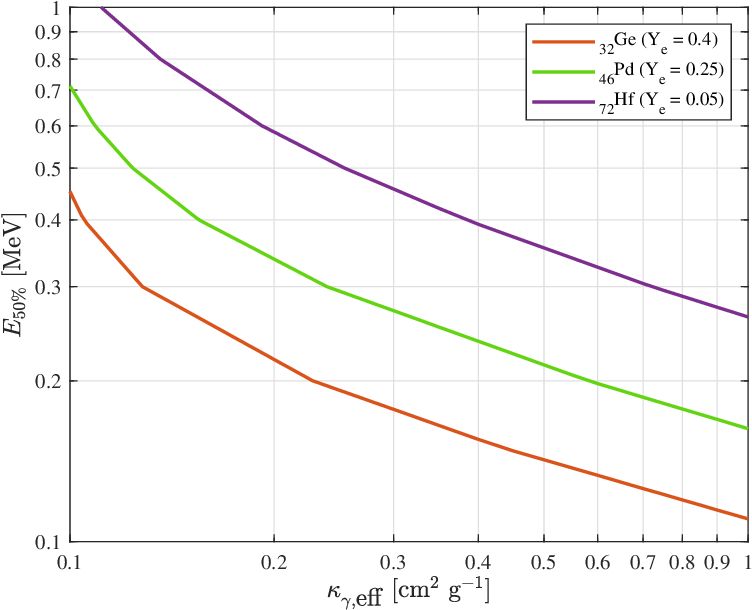}
 \caption{Upper limits on the median of the $\gamma$-ray energy-weighted spectrum, as given from eq. \eqref{eq:chernoff_kappa} with $x=0.5$, as a function of $\kappa_\geff$ and for different materials (in color). As in Fig. \ref{fig:kappa_gamma_general}, the opacities of $\{{}_{72}\text{Hf},{}_{46}\text{Pd},{}_{32}\text{Ge}\}$ represent well $\kappa_{\gamma,E}$ in NSM ejecta with initial $Y_e = \{0.05, 0.25, 0.4\}$ (and with $[\rho t^3]_{\text{KN}}=1$ and $s_0 = 20~\kb$).}
 \label{fig:chernoff_E_med}
\end{figure}

\section{Modifications to URILIGHT}
\label{ap:URILIGHT}
Here we list errors identified in the URILIGHT code, in the version published by \citet{wygoda_type_2019,blondin_standart_2022} (available at \url{https://www.dropbox.com/sh/kyg1z1xwi0298ru/AAAqzUMbr6AkoVfkSVIYChTLa?dl=0}), that were corrected in this work.
All modifications listed were applied to the \texttt{gammatransfer.f90} file. Note that (i) is relevant only for nickel-powered ejecta. The composition of NSM ejecta in the MC calculation was updated at each time according to \textit{SkyNet} results for the composition (see \S~\ref{subsec:comp_numerical}).
\begin{enumerate}
    \item The $\gamma$-ray deposition calculation did not take into account the effect of the changing composition, as ${}^{56}$Ni and ${}^{56}$Co decay, on the interaction cross sections of $\gamma$-rays. These were set based on the initial composition, leading to few \% error at late times (note, the change in composition is included in URILIGHT's UVOIR transport code).
    In the subroutine \texttt{track\_gamma}, these line should be added right after \texttt{sig\_pp=0.0d0}, before the cross section calculation:
\begin{verbatim}totatoms=atoms(ind_fe56,ind(i,j,k))
         +atoms(ind_co56,ind(i,j,k))
         +atoms(ind_ni56,ind(i,j,k))
call Ni56DecayChain(totatoms,teff(nt),
                    atoms(ind_ni56,ind(i,j,k)),
                    atoms(ind_co56,ind(i,j,k)),
                    atoms(ind_fe56,ind(i,j,k)))\end{verbatim}
    \item Pair production is by default turned off. However, two errors will lead to incorrect results if it is turned on. First, a typo in the cross section calculation, the line
    \begin{verbatim}total_pair_production(hnum,
                      atoms(1:Niso,ind(i,j,k)),
                      iso(1:Niso)%z,sig_pe)\end{verbatim}
should be replaced by
        \begin{verbatim}total_pair_production(hnum,
                      atoms(1:Niso,ind(i,j,k)),
                      iso(1:Niso)%z,sig_pp)\end{verbatim}
    Second, after PP event occurs $1-2q$ of the energy of the $\gamma$-rays packet is locally absorbed, where $q = \frac{1.022MeV}{E}$. However, URILIGHT absorbs only $\frac{1}{2}-q$ of the energy locally, thus
    \begin{verbatim}if (z.lt.(0.5d0+q)) then\end{verbatim}
    should be replaced by\begin{verbatim}if (z.lt.(2*q)) then\end{verbatim}
\end{enumerate}
\section{Modifications to Reaclib}
\label{ap:REACLIB}
The current version of \textit{SkyNet}, that is available online (\url{https://bitbucket.org/jlippuner/skynet/}), includes an old version of the REACLIB database. However, for several radionuclides with measured decay rates, this version either omits these rates or includes theoretical rates instead. Therefore, using \textit{SkyNet} with this version may lead to erroneous radioactive emission calculation (that was outlined in \S~\ref{subsec:KNe_model}).

As an example, the version supplied with \textit{SkyNet} omits the $\alpha$-decay of ${}^{212}$Po to ${}^{208}$Pb that has a half-life of 300ns (the reaction does exists as an inverse reaction to $\alpha$-capture on ${}^{208}$Pb, which occurs at a rate that is effectively zero, at low-$T$).
Thus, low-$Y_e$ cases lead to accumulation of ${}^{212}$Po, which does not decay \citep[this can be seen in the final abundance of many works that used \textit{SkyNet}, e.g.][which includes a fraction of mass at $A=212$, despite the fact that there are no known stable isotopes at this $A$]{lippuner_r-process_2015}. As a result, using ENDF data, that includes the $\alpha$-decay of this isotope, to calculate the radioactive emission leads to a very large and non-decaying (constant) activity of ${}^{212}$Po, that then leads to this reaction wrongly dominating $\Dot{Q}_{\rm dep}$.
This discrepancy between the databases can be identified from the fact this $\Dot{Q}_{\rm dep}$ would differ from the one calculated by \textit{SkyNet} (as the change in the mass of isotopes, $\Sigma_{\rm iso}\Dot{M}_{\rm iso}c^2$, see Appendix \ref{ap:Checks}), as in \textit{SkyNet} ${}^{212}$Po does not decay and thus has zero activity.

To correct such issues we first updated the version of REACLIB used to its latest version (from June 2021). In this version, many missing or theoretical decay rates were replaced by measured values (including the $\alpha$-decay of ${}^{212}$Po). However, we have identified several issues with this version, and corrected them in the following way:
\begin{enumerate}
    \item ${}^{48}$Ca$\rightarrow$${}^{48}$Sc: the rate (with half-life of 30 yr) was deleted, since ${}^{48}$Ca has half-life $>10^{19}$ ry.
    \item ${}^{50}$V$\rightarrow$${}^{50}$Ti: given half-life is 100 d instead of $>10^{17}$ yr, so this rate should be changed/removed. This was not changed, as it was recognized after the computations were made. Anyway, it is of very low impact, as the decay product ${}^{50}$Ti is anyway stable, and it is relevant only for very neutron-poor cases.
    \item ${}^{100}$Mo$\rightarrow$${}^{100}$Tc: the rate is correct, but the decay should be to ${}^{100}$Ru (double $\beta$-decay). This was not changed, as anyway the half-life of ${}^{100}$Mo is $>10^{18}$ yr.
    \item ${}^{123}$Te$\rightarrow$${}^{123}$Sb: rate changed to half-life of $\sim$$10^{17}$ yr (from the reference wc12) instead of half-life of 100sec (from mo97).
    \item ${}^{128}$Te$\rightarrow$${}^{128}$I: rate was incorrectly labeled as inverse reaction, with "v" label (hence ignored by \textit{SkyNet}), label was removed.
\end{enumerate}

\section{validity of the composition and radioactive decay calculations}
\label{ap:Checks}
As noted in \S~\ref{subsec:KNe_model}, we describe here the verification that (i) at times considered (>1 hour) the ejecta energy production is dominated by radionuclide which have experimental radioactive decay data (half-lives and spectra); (ii) that the effect of the wrong SF rates in \textit{SkyNet} is small.

We conduct the first test by comparing the energy production rate calculated by \textit{SkyNet}, as $\Sigma_{\rm iso}\Dot{M}_{\rm iso}c^2$ ($M_{\rm iso}$ is the mass per baryon of a specific isotope in the ejecta), to the total energy release rate (including neutrino energy) that we compute, using the detailed composition and ENDF data, when we only include isotopes with complete decay data (half-lives and spectra).
Both agree completely from $\sim$1 hour, showing that there are no substantial inconsistencies between the experimental decay data and the experimental and theoretical decay data in REACLIB (such as the conflict in the rate of ${}^{212}$Po at older versions of REACLIB, see Appendix \ref{ap:REACLIB}). This confirms that from $\sim$1 hour, the radioactive energy production is dominated by isotopes which have complete radioactive decay data, and $\gamma$-ray spectra in particular. Of course, this cannot rule out the possibility that we miss energy production from radionuclides with no experimental data and a decay rate underpredicted by REACLIB.
At earlier times, for low $Y_e$, \textit{SkyNet} predicts an energy rate larger by a factor of $\sim$2 compared to ours, indicating that isotopes without experimental data cannot be neglected at these times.
As noted in \S~\ref{subsec:KNe_model}, due to wrong SF rates, the energy production calculated by \textit{SkyNet} overestimates our calculation. After subtracting the contribution of SF from $\Sigma_{\rm iso}\Dot{M}_{\rm iso}c^2$, we found again an excellent agreement between the calculations.

To make sure that the wrong SF rates do not affect our results, we conducted \textit{SkyNet} calculations for which we replaced the SF rates file, with a short list containing only few experimental SF rates of isotopes, for which SF is the main decay channel and the half life is relevant to our timescales. We observed very small changes in our results, $\lesssim$1\% in $\kappa_\geff$, even at low $Y_e$. This is expected, as despite the wrongly large rate of SF, it influenced only a small fraction of the composition, and dominated the energy release due to its large Q-value.

\section{Examination of the $\gamma$-ray spectra obtained by Barnes et al. 2021}
\label{ap:Barnes}
Fig. \ref{fig:chernoff_barnes} shows the results of \citet{barnes_kilonovae_2021} for the energy-weighted cumulative distribution function (CDF; $P(E_{\gamma}\leq E)=\int_0^{E}{dE\phi_\gamma(E,t)}$), computed at $t=\text{1 d}$, for different nuclear mass models and $Y_e$. Also shown are our results, for similar ejecta parameters and using the FRDM nuclear mass model.
To illustrate that $\kappa_\geff \gg 0.03-0.05$\cmgr requires a $\gamma$-ray spectra that are much softer than seen in numeric simulations, we show lower bounds on $P(E_\gamma \leq E)$ (obtained from eq. \ref{eq:chernoff_kappa}) that must be surpassed for $\kappa_\geff = 1, 0.2$\cmgr to be possible. Clearly, $\kappa_\geff\sim1$\cmgr requires $\gamma$-ray spectra which are much softer than any simulated spectra.

\begin{figure}
\centering
 \includegraphics[width=\columnwidth]{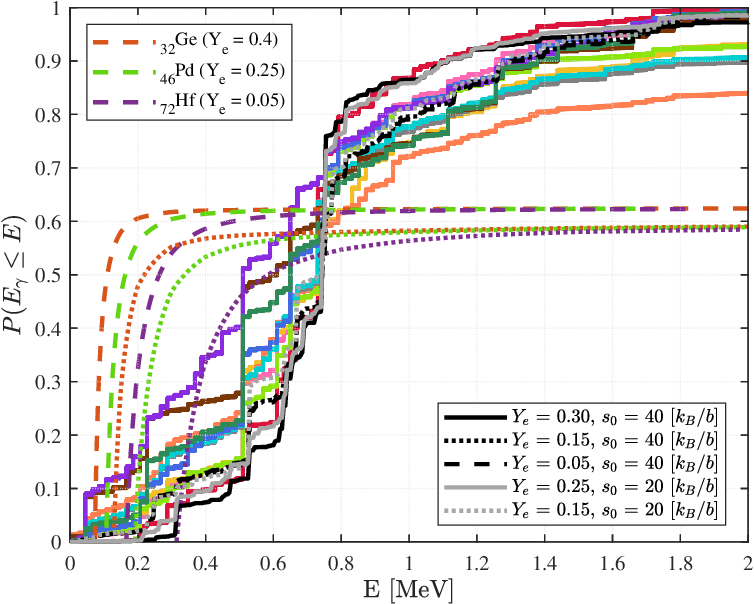}
 \caption{Colored solid lines show the $\gamma$-ray energy-weighted CDFs, $P(E_{\gamma}\leq E)=\int_0^{E}{dE\phi_\gamma(E,t)}$, obtained by \citet{barnes_kilonovae_2021}, at $t=\text{1 d}$ for $s_0 = 40\kb$, various low values of $Y_e$ and different nuclear mass model (adapted from their figure 2).
 Black and gray lines show our results for the CDFs, computed with the FRDM nuclear mass model for ejecta with $[\rho t^3]_{\text{KN}} = 2$ \citep[similar to the $\rho t^3$ value of the ejecta used in][]{barnes_kilonovae_2021}, $s_0 = 40 \kb$ (black) or $s_0 = 20 \kb$ (gray), and various $Y_e$ values as specified in the legend.
 In colored dashed and dotted lines we show lower bounds on $P(E_\gamma\leq E)$ for $\kappa_\geff = 1, 0.2$\cmgr, respectively, computed using eq. \eqref{eq:chernoff_kappa} - the CDF must surpass the lower bound at all energies for the given value of $\kappa_\geff$ to be possible. The lower bounds were computed for ejecta composed of ${}_{32}$Ge (orange), ${}_{46}$Pd (green) and ${}_{72}$Hf (purple). As in Fig. \ref{fig:kappa_gamma_general}, the opacities of these elements represent well the $\kappa_{\gamma,E}$ in NSM ejecta with $Y_e=0.4$, $Y_e=0.25$ and $Y_e=0.05$ cases, respectively (with $[\rho t^3]_{\text{KN}}=1$ and $s_0 = 20~\kb$).
 The red and pink lines of \citet{barnes_kilonovae_2021}, which were obtained for the FRDM nuclear mass model and $Y_e=0.28,0.16$, respectively, agree well with our results for similar parameter values (a better agreement at low energies is obtained using our results for $s_0 = 20~\kb$ and not $s_0 = 40~\kb$, which may point at possible deviations between the used codes).
 Our result for $Y_e = 0.05$ is given to show that it nearly coincides with that for $Y_e = 0.15$, in accordance with fission cycling, discussed in \S~\ref{subsec:KNe_kappa_geff}.}
 \label{fig:chernoff_barnes}
\end{figure}


\bsp	
\label{lastpage}
\end{document}